\documentclass[12pt,preprint]{aastex}

\shorttitle{Line-Driven Accretion Disk Winds. III.}
\shortauthors{Pereyra & Kallman}

\begin{document}

\title{Hydrodynamic Models of Line-Driven Accretion Disk Winds III:
       Local Ionization Equilibrium}

\author{Nicolas Antonio Pereyra\altaffilmark{1}}
\affil{University of Pittsburgh, Department of Physics and Astronomy,
       3941 O'Hara ST, Pittsburgh, PA 15260}
\email{pereyra@bruno.phyast.pitt.edu}

\and

\author{Timothy R. Kallman}
\affil{NASA/GSFC, Laboratory for High Energy Astrophysics,
       Code 662, Greenbelt, MD 20771}
\email{tim@xstar.gsfc.nasa.gov}

\altaffiltext{1}{Universidad Mar\'\i tima del Caribe,
                 Departamento de Ciencias B\'asicas,
                 Catia la Mar, Venezuela}

\begin{abstract}
We present time-dependent numerical hydrodynamic models of
line-driven accretion disk winds in cataclysmic variable systems and
calculate wind mass-loss rates and terminal velocities.
The models are 2.5-dimensional,
include an energy balance condition with radiative
heating and cooling processes,
and includes local ionization equilibrium introducing time dependence
and spatial dependence on the line radiation force parameters.
The radiation field is assumed to originate in an optically
thick accretion disk.
Wind ion populations are calculated under the assumption that
local ionization equilibrium is determined by
photoionization and radiative recombination,
similar to a photoionized nebula.
We find a steady wind flowing from the
accretion disk.
Radiative heating tends to maintain the temperature in the
higher density wind regions near the disk surface,
rather than cooling adiabatically.
For a disk luminosity $L_{disk}=L_{\sun}$,
white dwarf mass $M_{wd}=0.6 M_{\sun}$,
and white dwarf radii $R_{wd}=0.01 R_{\sun}$,
we obtain a wind mass-loss rate of
$\dot M_{wind}=4 \times 10^{-12} M_{\sun} \, {\rm yr}^{-1}$,
and a terminal velocity of $\sim 3000 {\rm \, km \, s}^{-1}$.
These results confirm the general velocity and density
structures found in our earlier constant ionization equilibrium
adiabatic CV wind models.
Further we establish here 2.5D numerical models that can be extended
to QSO/AGN winds where the local ionization equilibrium will play a
crucial role in the overall dynamics.
\end{abstract}


\section{Introduction}

Ultraviolet observations of Cataclysmic Variables (CVs) show that
virtually all nonmagnetic CVs with high mass accretion rates
($\gtrsim 4\times 10^{-10} M_{\sun} \, {\rm yr}^{-1}$)
exhibit P~Cygni profiles in the resonance lines
\citep[e.g.][]
{cor82,gre82,gui82,kla82,vit93,pri95,fri97,gan97,kni97,pri00}.
These profiles consist of blue-shifted absorption and
redshifted  emission,
with the emission most prominent in high inclination systems
\citep[e.g.][]{hut80,kra81,hol82,cor85,mas95,che97},
and the absorption appearing at low and intermediate inclinations.
The most straightforward explanation for these observations is in
terms of a wind originating from the accretion disk.
The wind is driven by radiation pressure from the disk UV continuum
absorbed by line transitions,
analogous to the radiation driven winds in early type stars.
Support for this scenario came from \citet{dre87},
who calculated theoretical \ion{C}{4} 1549 \AA \  line-profiles from
one-dimensional kinematic CV wind models,
for a wide variety of wind and disk parameters.
\citet{dre87} found that the  dependence on the form of the UV
resonance lines with inclination angle could be accounted for by
a bipolar wind emerging from an accretion disk.

One-dimensional dynamical models for line-driven winds (LDW) in CVs
were developed independently by \citet{vit88} and \citet{kal88}.
Further efforts came through two-dimensional kinematic modeling
\citep{shl93,kni95},
which succeeded in showing consistency between the assumed polar
geometry of a disk wind and observed profiles;
but such models were unable to self-consistently calculate the
wind dynamics including the effects of rotation and the anisotropic
disk radiation field.

Two-dimensional disk wind dynamical models were developed by
\citet{ick80},
but these did not take into account the radiation pressure due to line
absorption.
For a typical white dwarf mass of $0.6M_{\sun}$ \citep{lei80,sil01}
the disk luminosity required to produce such a wind without line
radiation pressure
(assuming radiation pressure due to continuum scattering only)
would require a luminosity of $\sim 10^5 L_{\sun}$ \citep{ick80},
several magnitudes above the observed luminosity of CVs which vary
between $0.01L_{\sun}$ and $10 L_{\sun}$ \citep{pat84}.
\citet{ick80} did find that,
for sufficiently high disk luminosities,
a biconical disk wind would form.
\citet{ick81} suggested that biconical winds were a general property
of accretion disk winds independent of the wind driving mechanism.

Two-dimensional dynamical models of isothermal line-driven
disk winds were presented by us \citet[hereafter Paper~1]{per97a}.
Results from Paper~1 showed,
in analogy with line-driven winds from early type stars,
that terminal velocities are approximately independent
of the luminosity of the disk, although increments in luminosity
produce increments in mass-loss rate, 
and that rotational forces are important in the study of winds from
accretion disks.
They cause the velocity streamlines to collide,
which reduce the speed and increase the density
of the wind producing an enhanced density region.
The highest absorption occurs in the enhanced density region where
density is increased relative to a spherically diverging wind with
the same mass loss rate and the velocity is roughly half the terminal
velocity.
This density increase is necessary in order to produce at least
marginally optically thick lines.

We developed \citep[hereafter Paper~2]{per97b,per00}
an adiabatic wind model
(rather than isothermal),
driven by a standard accretion disk
\citep{sha73,lyn74}
rather than an isothermal disk.
The hydrodynamic models developed of line-driven accretion
disk winds (LDADWs) in
CVs included the radial structure of an optically thick accretion
disk with the corresponding radiation fields and surface temperature
distributions.
The corresponding energy conservation equation was
also implemented, 
including the adiabatic heating and cooling effects due to
compression and expansion.
From the computational models,
assuming single scattering and constant ionization,
we calculated theoretical line profiles for the 
\ion{C}{4} 1550~\AA \ line,
and found that the line profiles obtained were
consistent with observations in their general form and strong
dependence with inclination angle.

\citet{pro98} also developed two-dimensional
models for these systems,
obtaining similar results to the models presented here.
They assumed an isothermal wind and constant values for the line
radiation parameters throughout the wind.
They found similar wind velocities and wind mass loss rates
as we do here,
although they assumed disk luminosities about an order
of magnitude greater.
A difference we find with the results of the models presented by
\citet{pro98} is that they obtain unsteady flows
characterized by large amplitude fluctuations in velocity and density,
while for our previous models
(Paper~1; Pereyra~1997; Paper~2)
and in the models presented here we find a steady flow.
We note that these large fluctuations
(rather than the steady flow we find)
could account for the difference in luminosity required by
the Proga models,
with respect to our models,
to obtain similar wind mass loss rates.
One possibility suggested by \citet{pro98} was that the spatial
resolution of the models we presented in Paper~1 was too low to
adequately resolve the wind.
In paper~2 we increased significantly the spatial resolution of
our models and obtained similar steady results when compared to
our earlier spatial resolutions used in Paper~1;
thus showing that differences in spatial resolution of the
models was not generating the difference between our steady disk wind
models and the ``intrinsically unsteady'' disk winds reported by
\citet{pro98}.
Another possibility is that the difference in the unsteady flows 
of the Proga models with respect to the steady flows we find
is due to the boundary condition treatment in the Proga models,
which appears to lead to instabilities at the base of the wind and in
turn to the strong fluctuations reported.

It is well known that for early type stars,
when a Sobolev treatment is used for the line radiation
pressure (as we have done in our models and as has been
applied by \citet{pro98} for accretion disk winds in CVs),
a steady wind solution is obtained \citep[e.g.][]{owo99}.
The treatment of boundary conditions in numerical models for
line driven winds in early type stars was discussed by
\citet{owo94} and \citet{cra96}.
They find that the treatment of the lower (base of wind)
boundary condition is ``somewhat problematic''.
Further they find that if they used an arbitrarily high
density,
similar to the boundary conditions used by \citet{pro98},
they would find oscillations at the base of the wind in
both density and velocity.
In both papers,
in addition to other details in the boundary
condition treatment \citep{owo94,cra96},
they lowered the density at the inner boundary until
they found steady conditions at the base and were
simultaneously able to resolve acceleration in the subsonic
region.
Therefore  a conclusion in the area of numerical modeling that we
derive is that by varying the treatment of boundary
conditions at the base of a line driven wind,
one can obtain strong fluctuations in density and velocity
which are not necessarily physically intrinsic but are rather
a consequence of the numerical details of the model.

\citet{fel99a} and \citet{fel99b} developed two-dimensional
stationary models for CV disk winds where they solve the dynamics
under the assumption that the wind streamlines lie on straight cones.
The values of wind mass loss rates obtained through
the models presented by \citet{fel99b} for an isothermal disk
(equation~[11] of their paper) are consistent with those found in
Paper~1.
In addition \citet{fel99b} find that in an isothermal disk
(equations~[10]-[11] of their paper)
the wind mass loss rate scales as
$\dot M_{wind} \propto L_{disk}^{1/\alpha}$,
where $L_{disk}$ is the disk luminosity and $\alpha$ is line
force multiplier parameter.
This is the same scaling law reported in Paper~1.
Further, the biconical geometry
and wind tilt angle in the models of \citet{fel99b}
are roughly consistent with the  velocity structures we find through
full 2.5D hydrodynamic modeling of our earlier models
(Paper~1; Pereyra~1997; Paper~2)
and in the models presented here.
We further note that Feldmeier et~al. suggest that detailed
non-LTE calculations of the line-force multiplier may be important
in the modeling of CV disk winds.
This work is an additional step in that direction.

In spite of these encouraging results,
a more realistic model which would include a detailed
calculation of the UV spectrum from dynamically calculated
density and velocity fields is desirable in order to make
specific comparisons with observations.
For instance,
a maximum flux  greater than $1.7$ times the continuum
has been reported for the emission component of P~Cygni line
profiles of \ion{C}{4} 1550\AA \ from some cataclysmic
variables \citep[e.g.][]{cor82},
while from our previous models
(Pereyra~1997; Paper~2)
we find that the emission maximum of the \ion{C}{4} lines stays below
1.3 times the continuum flux for those angles where the absorption
component is observed.
This discrepancy may be due to some of the simplifications in our
previous models or it may indicate a \ion{C}{4} line-emitting region
other than the disk wind itself.
Furthermore,
in recent observational results,
\citet{fro01} find relatively narrow absorption lines
($\leq 700 \, {\rm km \, s}^{-1}$) in the FUV spectra of the
cataclysmic variable U~Gem.
As \citet{fro01} have indicated,
such lines cannot arise from the disk photosphere.
In their work they suggest the possibility of an outer disk
chromosphere to account for these lines,
but it is also possible that some of these narrow FUV lines could be
accounted for by an accretion disk wind.

An additional motivation for this work is that the models
developed here can also be extended to the study of LDADWs in
Quasars (QSOs),
where the local ionization equilibrium will
play an important in the overall dynamics \citep{mur95,hil02}.
Evidence for winds in QSOs is found in the broad absorption lines
(BALs) observed in approximately 10\% of the QSOs
\citep[e.g.][]{wey97}.
If the wind in QSOs is generated from an accretion disk,
then the presence of BALs in only a fraction of QSOs
can be accounted for as a viewing-angle effect
\citep{tur84};
similar to the case of CVs where the accretion disk wind produces
P~Cygni profiles when observed ``face-on'' and emission lines
(without absorption components) when seen ``edge-on''
\citep{per00}.
The x-ray luminosity in QSOs are generally comparable to the
UV/optical luminosities \citep[e.g.][]{tan79,gri80,mus93,lao97,geo00}.
\citet{dre84} found that such high x-ray luminosities would ionize
the wind to a point where the populations of ions responsible for
the BAL observed in QSOs would be too low to produce observable
absorption lines.
\citet{mur95} found that,
with an appropriate x-ray shielding mechanism,
LDADWs can produce wind densities and velocities consistent
with BALs observed in QSOs.
Strong changes in the ionization balance can produce observable
effects in line formation in QSOs;
as well as significant changes in the line radiation force parameters
which will play an important role in the overall wind dynamics
\citep{hil02}.

In our previous models
(Pereyra~1997; Paper~2)
we had assumed constant ionization equilibrium throughout the wind and
single scattering in the line profile calculations.
In this work we extend our previous models to include radiative heating
and cooling and ionization balance in the wind,
and a calculation of the line radiation force parameters at each
point in space at each time step.
Our results are similar to those of Paper~2,
and therefore act to confirm the general results found in our
previous simpler models.
They are a further step towards a model capable of self-consistently
producing theoretical spectrum which may be compared in detail with
observations;
they can also be extended to the study of LDADWs in QSOs/AGN,
where the local ionization equilibrium will
play a crucial role in the overall dynamics \citep{mur95,hil02}.

In \S\ref{sec_adisk} we discuss the radial structure of the
accretion disk and the radiation field as implemented in our
models.
In \S\ref{sec_ioneq} we present the ionization balance calculations
applied in this work.
We derive the expression used for the treatment of the
line radiation pressure in \S\ref{sec_rf}.
The treatment of radiative heating and cooling in this work is
discussed in \S\ref{sec_rhc}.
In \S\ref{sec_hyd} we present and discuss the hydrodynamic
calculations. 
The results of the models are presented and discussed in
\S\ref{sec_res}.
We present a summary and the conclusions of this work in
\S\ref{sec_sum}.

\section{Accretion Disk}
\label{sec_adisk}

A key ingredient in the models presented in this work is the
radiation field emerging from the accretion disk.
The boundary layer where the disk intercepts the white dwarf and the
white dwarf photosphere may also contribute to the radiation field.
The white dwarf by itself
(without considering the effects of the accreting mass)
will have a luminosity of the order of $0.01L_{\sun}$ \citep{lei80},
while the accretion disk will have typical luminosities of the
order of $L_{\sun}$ \citep{war87}.
As we indicated in Paper~2,
there is still uncertainty concerning the existence and spectrum
of the boundary layer
\citep{vrt94,mau95,lon96,szk99,mau00,fro01}.
In the models presented here we have chosen not to explore the
boundary layer issue in depth,
but instead to consider the non-boundary layer scenario.
We assume in this work that the radiation field is generated by
the accretion disk alone,
neglecting any boundary layer or white dwarf radiation.
We assume here the standard accretion disk
\citep{sha73,lyn74}.
In paper~2 we discussed some of the observational evidence for
the existence of an accretion disk in CVs.
We note that in this work we study the winds originating in
nonmagnetic CVs in which an accretion disk is formed.

In the standard accretion disk model,
the disk is assumed to be in a steady state.
Shear stresses transport angular momentum outwards as the
material of the gas spirals inwards.
Conservation of angular momentum leads to the following expression:
\begin{equation}
W 2\pi r^2 - \dot{M}_a \omega r^2 = C \, ({\rm constant})
\end{equation}
where $r$ is the radius,
$\dot{M}_a$ is the mass accretion rate,
$\omega$ is the corresponding angular velocity.
and $W$ is defined by:
\begin{equation}
W \equiv \int\limits_{-z_0}^{z_0} w_{r \phi} \, dz
\end{equation}
where $z_0$ is the half thickness of the disk and
$w_{r \phi}$ is shear stress between adjacent layers.
With the additional assumption that the shear stresses
are negligible in the inner disk radius:
\begin{equation}
\label{equ_stress}
W = {\dot{M}_a \over 2\pi r^2} ( \omega r^2 - \omega_0 r_0^2)
\end{equation}
where $\omega_0$ is the angular velocity at the inner disk
radius $r_0$. In CVs the inner disk radius corresponds to
the white dwarf radius $R_{wd}$,
i.e. the accretion disk extends from the inner Lagrangian point
of the binary system down to the white dwarf surface.

Taking into account the work done by shear stresses,
and assuming that, as the mass accretes inwards,
the gravitational energy lost is emitted locally:
\begin{equation}
\label{equ_qdisk_gen}
Q = {1 \over 4\pi r} \; {d \over dr}
           \left[
             \dot{M}_a \left(  {\omega^2  r^2 \over 2}
                         - {GM \over r}
                       \right)
             - W 2\pi \omega r^2
           \right]
\end{equation}
where $Q$ is the radiated energy per area of the disk surface.
Assuming that the disk material follows Keplerian orbits, i.e.
\begin{equation}
\omega = \left( {G M_{wd} \over r^3} \right)^{1/2}
\end{equation}
where $G$ is the gravitational constant and $M_{wd}$ is the
white dwarf mass;
from equations~(\ref{equ_stress})-(\ref{equ_qdisk_gen}) one finds:
\begin{equation}
\label{equ_qdisk}
Q = {3 \dot M_{accr} G M_{wd} \over 8 \pi r^3}
    \left[1 - \left({r_o \over r} \right)^{1/2} \right]
\end{equation}

The function $Q(r)$ (equation~[\ref{equ_qdisk}]) was originally
derived by \citet{sha73} for binary systems with an accretion disk
about a black hole.
\citet{pop95} developed detailed models of boundary layers
and accretion disks in cataclysmic variables and found that their
models supported the same function $Q(r)$ for the accretion disk
in CVs.

Further,
assuming that the disk is emitting locally as a blackbody,
the radial temperature distribution of the disk will be
given by:
\begin{equation}
\label{equ_tdisk}
T = \left\{ {3 \dot M_{accr} G M_{wd} \over 8 \pi r^3 \sigma_s}
    \left[1 - \left({r_o \over r} \right)^{1/2} \right] \right\}^{1/4}
\end{equation}
where $\sigma_s$ is the Stefan-Boltzmann constant. 
Equation~(\ref{equ_qdisk}) is implemented in our model in the
calculation of radiation flux throughout the wind,
and consequently in the calculation of the continuum and line radiation
pressure force.
We assume local blackbody radiation.
In our model we also assume that the temperature at the base of the
wind is equal to the disk surface temperature for each radius.
We wish to note here that the blackbody assumption of
equation~\ref{equ_tdisk} will not strictly hold.
Radiative diffusion of the energy from the inner parts of the
disk to the disk photosphere will produce deviations from the
blackbody assumption,
limb darkening effects will produce deviations from the projected
area in the angular dependence of the emitted flux,
line scattering and electron scattering will produce deviations
from the blackbody emission,
the relatively high velocity gradients in the rotational shears as
well as the wind velocity gradients will affect the line scattering;
thus a detailed calculation of the spectrum emitted from the
accretion disk would requires a 2.5D radiative transfer calculation of
the disk atmosphere coupled with the corresponding velocity gradients.

Early attempts to calculate deviations of the actual disk spectrum
from the local-blackbody assumption,
applied locally spectra from stellar atmosphere codes
\citep{her79, kip79, kip80, may80, kol84}.
More recently,
local plane parallel radiative transfer models implementing
disk vertical structure have been developed
\citep{dia96,hub97,hub98,wad98,hub00}.
However,
to our knowledge a self-consistent 2.5D radiative transfer
model coupled with velocity shears of an accretion disk
atmosphere has not yet been developed.
Also, there still remains uncertainties regarding the variation
of viscosity in the disk which in turn effects the vertical structure.
Thus,
due to its simplicity,
the assumption of local blackbody spectrum has often been used
to model accretion disk and their spectra
\citep[e.g.][]{shl93,kni95,pro98,per00}.
In future work we shall consider the radiative transfer effects
that deviate the disk spectrum from an ideal local-blackbody,
and attempt to couple them with hydrodynamic calculations.

It is generally accepted that the mechanism for producing the shear
stress in accretion disks is magnetic turbulence generated
by the Balbus-Hawley magnetorotational instability (MRI) \citep{bal91}.
Three-dimensional magnetohydrodynamic simulations
\citep[e.g.][]{arm98,mac00,haw01a,haw01b}
suggest that the standard assumptions
(e.g. Keplerian orbits; zero stress at inner radii)
may not hold and that therefore corrections to
equation~(\ref{equ_qdisk}) may arise.
We believe that the standard Shakura-Sunyaev disk includes much
of the physics involved in accretion disks, such as:
gravity from a large central mass determining the
average velocity fields in the disk,
angular momentum transport,
conversion of gravitational energy to radiative emission,
and monotonic increase of temperature with decreasing radius
(except of radius close to the inner disk radii) and local
blackbody emission which accounts for the relatively ``flat'' 
continuum spectrum observed in systems where accretion
disk are inferred.
We also believe that it is important to explore ``simple'' models
where the number of free parameters is small.
More realistic disk models and radiation fields will be
introduced at a later date.

\section{Ionization Equilibrium and Line Opacity}
\label{sec_ioneq}

The main wind driving mechanism in our model is the line radiation
force generated as the disk emission encounters ionized gas above
the disk.
The wind material is thus driven from the disk photosphere achieving
velocities of the order of a few thousand of ${\rm km \, s}^{-1}$.
The line radiation force will depend on the available line opacity
distribution of the wind material,
which in turn is a direct function of the ion populations.
Therefore the ion populations become an important ingredient in our
models.
In our previous work (Paper~1, Pereyra~1997, Paper~2),
in an approach similar to that presented by
\citet{cas75} for OB star winds,
we had assumed constant ionization equilibrium throughout the wind.
Further we had assumed that the ion populations were similar to
those of early type stars.
These two assumptions allowed us to apply typical OB star line
radiation force parameters \citep{abb82} throughout the wind,
significantly simplifying the numerical treatment.
In this work we include the local ion population calculations
simultaneously with the hydrodynamics as we describe below.

The ion populations of the wind material is initially at local
thermodynamic equilibrium (LTE),
since the wind starts at the disk photosphere.
As the wind material rises,
the ion populations are no longer in LTE.
The material is irradiated by the disk,
which is a non-Plank source
(equations [\ref{equ_qdisk}]-[\ref{equ_tdisk}]).
Due to the relatively low wind density,
the ion populations above the disk are largely determined by 
photoionization and radiative recombination processes,
which in turn are dependent on the incoming disk radiation.
It is difficult to model the collisional and radiative processes
which determine the ion populations,
varying in space and time,
while simultaneously solving the hydrodynamics of the system.
In this work we assume that the ion populations in the disk wind are
determined by photoionization and radiative recombination processes,
similar to a photoionized nebula irradiated by an accretion
disk.
To illustrate the validity of this approximation we shall estimate the
ratio between the radiative photoionization rate $R_{ij}$ and the
collisional ionization rate $C_{ij}$.
Under the condition that $h\nu_o \gg kT_R$ and $h \nu_o \gg kT_e$,
where $h$ is Plank's constant,
$h\nu_o$ is the minimum energy required to ionize an atom from state
$i$ to state $j$,
$k$ is Boltzmann's constant,
$T_R$ is the temperature of the ionizing radiation,
and $T_e$ the electron temperature,
the ratio can be approximated by \citep{mih78}:
\begin{equation}
\label{equ_mihalas}
{R_{ij} \over C_{ij}} \approx
{ 4 (2 \pi^3 k)^{1 \over 2} h \nu_o^3 \over
  3 m_e^{1 \over 2} e^2 c^3}
\left( {W T_R \over n_e T_e^{1 \over 2}} \right)
\exp \left[ h \nu_o
\left( {1 \over k T_e} - {1 \over k T_R} \right) \right]
\end{equation}
where $m_e$ is the electron mass,
$e$ the electron charge,
$c$ the speed of light,
$W$ the dilution factor of the radiation from its source,
and $n_e$ is the electron density.
Taking for the CV case: $n_e \sim 10^8$ (this density is a typical
value for the CV disk wind models we presented in Paper~2),
$T_R \sim T_e \sim 30,000 \, {\rm K} \;$,
$W \sim1/2$,
and $h \nu_o \sim 13.6 \, {\rm ev}$,
from equation~(\ref{equ_mihalas}) we find:
\begin{equation}
{R_{ij} \over C_{ij}} \sim 1.4 \times 10^8 \; .
\end{equation}

As we have indicated above,
in order to calculate the force due to line scattering,
one in principle must consider the contribution to all lines
which in turn will depend on the ion populations at a particular point
in space in a particular time.
In the calculation of ion population and line opacity we follow
a similar scheme as that applied by \citet{abb82} for OB stars.
In this work we assume that photoionization and radiative
recombination processes determine the ion populations,
therefore the ionization balance in the wind is determined through
Saha's equation for a gas photoionized by an external source:
\begin{equation}
\label{equ_saha}
{N_{j+1} n_e \over N_j} =  2 W {U_{j+1} \over U_j}
                          \left( 2 \pi m_e k T_e \over h^2 \right)^{3/2}
                          {T_R \over T_e}
                          \exp\left( - {X_{Ij} \over k T_R} \right)
\end{equation}
where $N_j$ and $N_{j+1}$ are the atom densities at stage $j$ and
at the following ionization stage $j+1$,
$U_j$ and $U_{j+1}$ are the partition functions for the
corresponding ionization stages,
and $X_{Ij}$ is the ionization potential for the atom at stage $j$.
Equation~(\ref{equ_saha}) has been applied to several astrophysical
systems,
e.g. stellar winds \citep{abb85,sch90,luc93}.

In the work presented here,
we have assumed that the radiation is primarily coming
from the accretion disk which presents a temperature
dependent on radius (equation~[\ref{equ_tdisk}]).
Equation~(\ref{equ_saha}) assumes that the photoionization rates are
much greater than collisional ionization rates,
thus:
\begin{equation}
N_j I_j \approx N_{j+1} R_{j+1}
\end{equation}
where $I_j$ is photoionization rate for an atom at ionization
stage $j$, and $R_{j+1}$ is the radiative recombination rate for
the atom at ionization state  $j+1$.
Therefore, from equation~(\ref{equ_saha}):
\begin{equation}
I_j  = R_{j+1}   {2 W \over n_e} {U_{j+1} \over U_j}
                          \left( 2 \pi m_e k T_e \over h^2 \right)^{3/2}
                          {T_R \over T_e}
                          \exp\left( - {X_{Ij} \over k T_R} \right) \; .
\end{equation}
In the accretion disk,
the radiation is emitted at different temperatures
(equation~[\ref{equ_tdisk}]),
thus the total photoionization rate $I_{Tj}$ due to the disk radiation
must equal the sum of all the contributions of photoionization rates
due to the radiation emitted at each section of the disk surface:
\begin{equation}
I_{Tj} = R_{j+1} {2 \over n_e} {U_{j+1} \over U_j}
                           \left( 2 \pi m k T_e \over h^2 \right)^{3/2}
                           { 1 \over T_e}
                           \int_{disk}
                           { W(r,\phi) T_R(r)
                             \exp\left( - {X_{Ij} \over k T_R(r)}
                                     \right)} r \, d\phi \, dr
\end{equation}
where $W(r,\phi)$ is the dilution factor per source area,
given by:
\begin{equation}
W(r,\phi) = {1 \over 4 \pi} \>
            { {\hat z} \cdot
              (\vec r_p -\vec r) \over |\vec r_p - \vec r|} \>
            {1 \over | \vec r_p - \vec r|^2}
\end{equation}
i.e.:
\begin{equation}
\label{equ_df}
W(r,\phi) = {1 \over 4 \pi} \>
            {z \over  [R^2 - 2Rr\cos\phi + r^2 + z^2]^{3/2} }
\end{equation}
where $\vec r_p \equiv (R,0,z)$ is the position where the
dilution factor is evaluated and 
$\vec r \equiv (r \cos\phi, r\sin\phi, 0)$ is the radiation source
position on the disk.

Therefore,
under the assumption that the photoionization and radiative
recombination processes dominate over the collisional processes,
we have:
\begin{equation}
\label{equ_saha_dist}
{N_{j+1} n_e \over N_j} =  2 {U_{j+1} \over U_j}
                           \left( 2 \pi m k T_e \over h^2 \right)^{3/2}
                           { 1 \over T_e}
                           \int_{disk}
                           { W(r,\phi) T_R(r)
                             \exp\left( - {X_{Ij} \over k T_R(r)}
                                     \right)} r \, d\phi \, dr
\end{equation}
which is the equation that we implement in this work for
the calculation of ionization balance for the line radiation
force.

Once the ion populations at different stages is calculated for
each atom through the resolution of equation~(\ref{equ_saha_dist}),
the population for each energy level for a given ion stage of
a given atom is obtained through Boltzmann's equations:
\begin{equation}
\label{equ_boltzmann}
{N_{ij} \over N_{0j}} = {g_{ij} \over g_{0j}}
                        \exp\left(-{X_{ij} \over k T_e} \right)
\end{equation}
where $N_{ij}$ and $N_{0j}$ are the densities of a given atom
at a given ionization stage $j$ at energy level $i$ and at
the lowest energy level respectively,
$g_{ij}$ and $g_{0j}$ are the degeneracies of energy level $i$
and at the lowest energy level respectively,
and $X_{ij}$ is the difference in energy between level $i$
and the lowest energy level.
We wish to note here that the LTE excited population assumption in
equation~(\ref{equ_boltzmann}),
while allowing the local ionization equilibrium calculations
to stay numerically manageable,
will tend to be more accurate for the lower energy ion states which
are the most crucial for the line radiation force.
For early type stars \citet{abb82} found that roughly 90\% of the
line radiation force was due to lines arising from ground
or metastable states.

After the ion populations at different stages at different
energy levels for a given atom have been obtained,
the line opacity of a given line for a given atom is calculated
by the expression: 
\begin{equation}
\label{equ_lopacity}
\kappa_L = {\pi e^2 \over m_e c} \, g_{lj} f_{luj} \,
           {N_{lj}/g_{lj} - N_{uj}/g_{uj} \over \rho} \,
           {1 \over \Delta\nu_D}
\end{equation}
where $g_{lj}$ and $g_{uj}$ are the degeneracies of the lower energy
level $l$ and the upper energy level $u$ of ionization stage $j$ for
a given line of a given atom,
$f_{luj}$ is the absorption oscillator strength corresponding to
the transition between lower energy level $l$ and upper energy
level $u$ of ionization stage $j$ for a given line of a given atom,
$N_{lj}$ and $N_{uj}$ are the ion densities of
ionization stage $j$, lower energy level $l$, and upper
energy level $u$ respectively,
$\rho$ is the total mass density of the gas,
and $\Delta \nu_D$ is the Doppler width of the given line.
The Doppler width of the line is given by:
\begin{equation}
\Delta \nu_D = {v_{th} \over c} \nu_L
\end{equation}
where $v_{th}$ is the thermal velocity and the line frequency $\nu_L$
is given by:
\begin{equation}
\nu_L = {X_{uj}-X_{lj} \over h} \; .
\end{equation}

\section{Radiation Force}
\label{sec_rf}

Once the line opacity $\kappa_L$ has been calculated through the above
expressions for each atomic line at each ionization stage,
the line radiation force is given by:
\begin{equation}
\label{equ_lf_gen0}
\vec{f}_{rad}= \sum_{lines} {\kappa_L \Delta\nu_D \over c}
          \oint{ \left[ {1 \over \tau_L(\hat{n}) }
               \int_0^{\tau_L(\hat{n})}e^{-\tau^\prime}d\tau^\prime
                    \right]
            I(\hat{n}) \, \hat{n} \, d\Omega}
\end{equation}
where $I(\hat{n})$ is the radiation intensity, and $\tau_L(\hat{n})$
is the line optical depth.

The line optical depth $\tau_L(\hat{n})$ for a given line at a
given direction $\hat{n}$,
under the Sobolev approximation,
is given by \citep{ryb78}:
\begin{equation}
\tau_L({\hat n}) = {\rho \kappa_L v_{th} \over 
                     |\hat{n} \cdot \vec{\nabla}
                      (\hat{n} \cdot \vec{v})| } \; .
\end{equation}

As in Paper~1 and Paper~2,
we assume in this work that the velocity gradient is primarily in the
``$z$'' direction and approximate the line optical depth by:
\begin{equation}
\label{equ_od}
\tau_L({\hat n}) \approx {\rho \kappa_L v_{th} \over 
                          \left| {\hat n} \cdot {\hat z}
                          \, {dv_z / dz} \right| } \; .
\end{equation}

Integrating equation~(\ref{equ_lf_gen0}) in optical depth:
\begin{equation}
\vec{f}_{rad}= \sum_{lines} {\kappa_L \Delta\nu_D \over c}
          \oint{
            \left[ {1 - \exp \left(
              - \tau_L(\hat{n}) \right) \over \tau_L(\hat{n})
             }
                    \right]
            I(\hat{n}) \, \hat{n} \, d\Omega}
\end{equation}

therefore:
\begin{equation}
\label{equ_lf_gen1}
\vec{f}_{rad} = \oint \Bigg[ 
         {\sigma \int_0^\infty I(\hat{n}) d\nu \over c }
                              \times \sum_{lines}
 { I(\hat{n}) \Delta\nu_D  \over \int_0^\infty I(\hat{n}) d\nu }
               \left( { 1 - \exp\left(- \, {\kappa_L \over \sigma }
                       \left[ {\sigma \over \kappa_L} \tau_L(\hat{n})
                       \right]\right) \over
                       {\sigma \over \kappa_L} \tau_L(\hat{n}) }
                  \right) 
 \Bigg] \hat{n} \, d\Omega 
\end{equation}
where $\sigma$ is Thomson cross section per unit mass.

In their work on radiation driven winds in OB stars,
\citet{cas75} and \citet{abb82},
found that the sum over lines in equation~(\ref{equ_lf_gen1}) could
be approximated by the following expression:
\begin{equation}
\label{equ_fm_def}
\sum_{lines}
 { I(\hat{n}) \Delta\nu_D  \over \int_0^\infty I(\hat{n}) d\nu }
               \left( { 1 - \exp\left(- \, {\kappa_L \over \sigma }
                       \left[ {\sigma \over \kappa_L} \tau_L(\hat{n})
                       \right]\right) \over
                       \left[ {\sigma \over \kappa_L} \tau_L(\hat{n}) 
                       \right] }
                  \right) 
  \approx 
k'(\hat{n}) \left( {\sigma \over \kappa_L} \tau_L(\hat{n})
   \right)^{- \, \alpha'(\hat{n})}
\end{equation}
where the $k'(\hat{n})$ and $\alpha'(\hat{n})$ are constants for
a given direction $\hat{n}$.
Defining $t$:
\begin{equation}
\label{equ_t}
t = {\sigma \over \kappa_L} \tau_L(\hat{n}) 
\end{equation}
we have:
\begin{equation}
\label{equ_fm0}
\sum_{lines}
 { I(\hat{n}) \Delta\nu_D  \over \int_0^\infty I(\hat{n}) d\nu }
               \left( { 1 - \exp\left(- \, {\kappa_L \over \sigma } t
                       \right) \over t }
                  \right) 
  \approx 
k'(\hat{n}) t^{- \, \alpha'(\hat{n})} \; .
\end{equation}
From equation~(\ref{equ_fm0}) we obtain the $k'(\hat{n})$ and the
$\alpha'(\hat{n})$ in a form similar to the \citet{abb82}
calculations for OB stars.
We calculate the left hand side of equation~(\ref{equ_fm0}) varying the
$t$ parameter from $10^{-1}$ to $10^{-7}$,
using the line opacity $\kappa_L$ calculated in the form described
in the previous section for each line,
and using the disk emission described by
equations~(\ref{equ_qdisk})-(\ref{equ_tdisk}) to
calculate the intensities.
From these values,
the parameters $k'(\hat{n})$ and $\alpha'(\hat{n})$ are obtained
through linear regression by varying the $t$ parameter over two
order of magnitudes at a time within the fore mentioned interval,
and taking the highest values found for each parameter.
We note here that the selection of the highest values within
intervals of two orders magnitudes is somewhat arbitrary.
To our knowledge this is the first attempt to include the spatial and
time dependence of the line radiation
force parameters coupled consistently with the hydrodynamics of
accretion disk winds, and we believe that the process includes 
much of the physics involved in line driving from a distributed
radiation source:
line-scattering,
photon momentum generating force on gas,
dilution factor,
line optical depth in a moving media,
line-opacity distribution,
energy distribution in spectrum,
and ion populations.
This approach allows us to explore the effect of
variable ion populations on the disk wind rather
that assuming a constant ionization equilibrium as we
did in Paper~2.
In future work we shall develop a more accurate treatment for
the line radiation force parameters,
including a systematic determination of the appropriate intervals
in $t$ in which to derive the line force parameters,
as well as a study of the directional dependence of these parameters,
rather than assume them isotropic as we do in this work
(see equation~[\ref{equ_fm1}]).

From equation~(\ref{equ_lf_gen1}) and equation~(\ref{equ_fm0}), we have:
\begin{equation}
\vec{f}_{rad} \approx \oint \Bigg[ 
         {\sigma \int_0^\infty I(\hat{n}) d\nu \over c }
k'(\hat{n}) t^{- \, \alpha'(\hat{n})}
 \Bigg] \hat{n} \, d\Omega \; .
\end{equation}

As mentioned above,
in this work we use values of the force multiplier parameters
$k$ and $\alpha$ which vary in time and space but are independent of
direction.
As we see below this will allow us to implement the
numerical integrations on intensity $I(\hat{n})$ at each spatial
point of the computational grid before implementing
the hydrodynamic calculations rather than during the implementation
of the hydrodynamic calculations,
thus maintaining the computations manageable.

The values of $k$ and $\alpha$ used in this work are calculated
through the expression:
\begin{equation}
\label{equ_fm1}
k \, t^{- \alpha} \approx { \left| \oint \Bigg[ 
         \int_0^\infty I(\hat{n}) d\nu
         k'(\hat{n}) t^{- \, \alpha'(\hat{n})}
         \Bigg] \hat{n} \, d\Omega  \right|
         \over
         \left| \oint \left[ \, {\int_0^\infty I(\hat{n}) d\nu} \,
          \right] \hat{n} \, d\Omega \, \right|} \; .
\end{equation}
We calculate the right hand side of equation~(\ref{equ_fm1}) varying
the $t$ parameter from $10^{-1}$ to $10^{-7}$,
using the values of $k'(\hat{n})$ and $\alpha'(\hat{n})$ calculated
through equation~(\ref{equ_fm0}) and using the disk emission described
by equations~(\ref{equ_qdisk})-(\ref{equ_tdisk}) to calculate the
intensities.
From these values,
the parameters $k$ and $\alpha$ are obtained through linear regression.
We note here that the angle averaging is done in an ad hoc manner in
order to reduce computational complexity.
As we indicated above,
we believe that this approach includes much of the physics of line
driving within an accretion disk wind.
In future work, we shall develop a more systematic treatment of
the line radiation force parameters that will include
a study of the directional dependence of these parameters,
rather than assume averaged angle values as we do here.

With these values of $k$ and $\alpha$,
we approximate the line radiation force by:
\begin{equation}
\vec{f}_{rad} \approx \oint \Bigg[ 
         {\sigma \int_0^\infty I(\hat{n}) d\nu \over c }
k \, t^{- \; \alpha}
 \Bigg] \hat{n} \, d\Omega \; .
\end{equation}

Therefore, from equation~(\ref{equ_t}):
\begin{equation}
\vec{f}_{rad} = \oint \left[ \,
         {\sigma \int_0^\infty I(\hat{n}) d\nu \over c } \,
 k \left( {\sigma \over \kappa_L} \tau_L(\hat{n})
                  \right)^{- \, \alpha} \,
   \right] \hat{n} \, d\Omega \; .
\end{equation}

Substituting equation~(\ref{equ_od}):
\begin{equation}
\vec{f}_{rad} = \oint \left[ \,
         {\sigma \int_0^\infty I(\hat{n}) d\nu \over c } \,
 k \left( { 1 \over \rho \sigma v_{th}}
           \left|\hat{n} \cdot \hat{z} \, {dv_z \over dz}
           \right| \, \right)^\alpha \,
   \right] \hat{n} \, d\Omega \; .
\end{equation}

Thus:
\begin{equation}
\label{equ_lf_spc0}
\vec{f}_{rad} = {\sigma \over c} \,
 k \left( { 1 \over \rho \sigma v_{th}} \,
      \left| {dv_z \over dz} \right| \, \right)^\alpha
          \oint \left[ \,
           \left|\hat{n} \cdot \hat{z} \right|^\alpha
         {\int_0^\infty I(\hat{n}) d\nu }
   \right] \hat{n} \, d\Omega \; .
\end{equation}

We wish to note that the $k$ and $\alpha$ parameters in the above
equation depend on the line opacity distribution
(equations~[\ref{equ_fm_def}]-[\ref{equ_fm1}]),
which in turn depends on density and temperature
(equations~[\ref{equ_df}]-[\ref{equ_lopacity}]).
Since the density and temperature will vary in space and time,
the $k$ and $\alpha$ parameters will also depend on space and time.
Also,
we wish to note here that in this work we assume azimuthal symmetry
and thus have only two independent spatial coordinates
``$r$'' and ``$z$''.

In equation~(\ref{equ_lf_spc0}) we substitute the $\alpha$ parameter
inside the integral by a fiducial value of $0.7$,
which is the value we assumed in our previous models presented
in Paper~1 and Paper~2.
Varying the number density from $10^6 \, {\rm cm}^{-3}$ to
$10^{10} \, {\rm cm}^{-3}$ at spatial points throughout our
computational grid,
and applying the values of disk luminosity
$L_{disk} = L_{\sun}$ and white dwarf radii of
$R_{wd} = 0.01 R_{\sun}$ which we use throughout this work,
by the way of equation~(\ref{equ_fm1}) we find values for the $\alpha$
parameter which vary from $0.55$ to $0.85$.
We do not expect that the application of the fiducial
value for $\alpha$ inside the integral of equation~(\ref{equ_lf_spc0})
will produce a significant difference in the density and
velocity wind structures,
and further it will allow the evaluation of the integral at all spatial
grid points before solving the hydrodynamic equations described 
in \S\ref{sec_hyd},
maintaining the computations manageable.
Thus,
\begin{equation}
\label{equ_lf_spc1}
\vec{f}_{rad} = {\sigma \over c}
 k(r,z,t) \left( { 1 \over \rho \sigma v_{th}} \,
      \left| {dv_z \over dz} \right| \, \right)^{\alpha(r,z,t)}
          \oint \left[ \,
           \left|\hat{n} \cdot \hat{z} \right|^{0.7}
         {\int_0^\infty I(\hat{n}) d\nu }
   \right] \hat{n} \, d\Omega \; .
\end{equation}

We wish to also note that although equation~(\ref{equ_lf_spc1})
takes into account the line radiation pressure in the ``$r$''
direction,
we have included only the derivative in the ``$z$'' direction
of the ``$v_z$'' component.
As we found in Paper~2,
the accretion disk winds in CVs tend to flow in the direction
perpendicular to the disk,
particularly in the region near the surface of the disk,
thus justifying the assumption that the derivatives of the ``$v_r$''
component are negligible.
In this work we also assume negligible the radial derivatives of the
``$v_z$'' component and the vertical and radial derivatives of the
``$v_\phi$'' component in the calculation of the line radiation force.
We note here that the disk wind streamlines are ``helical'' in nature;
the wind flow starts out at the base of the wind with a
non-zero ``$v_\phi$'' component due to disk rotation.
The ``$v_\phi$'' component of the velocity fields is calculated
throughout our hydrodynamic calculations under the assumption of
azimuthal symmetry (see Appendix).
We note that even under azimuthal symmetry,
line radiation force may have non-zero components along the
the ``$\phi$'' direction;
e.g., \citet{gay00} for the case of OB stars,
assuming azimuthal symmetry,
showed that the $\phi$ component of the line radiation force,
although presenting modest modifications in the overall wind dynamics,
could lead to observable effects.
Even though,
on one hand,
the inclusion of the additional terms in the velocity gradient for
the line radiation pressure in the accretion disk wind
may possibly produce interesting results,
on the other we do not believe that it will significantly affect the
overall results currently being implemented.
In future work we shall include the effects of the derivatives along
both the ``$z$'' and ``$r$'' directions for the different velocity
components in the line radiation force,
rather than assume ``$dv_z/dz$'' as the dominant term as we have done
in this work and in Paper~1 and Paper~2.

Defining $\vec S(r,z)$:
\begin{equation}
\label{equ_svec}
\vec S(r,z) = \oint
              \left[ \, \left|\hat{n} \cdot \hat{z} \right|^{0.7}
              {\int_0^\infty I(\hat{n}) d\nu} \,
              \right] \hat{n} \, d\Omega \; .
\end{equation}

Taking the source of radiation to be the accretion disk,
we have:
\begin{equation}
\label{equ_sz}
S_z(r,z) = \int_{r_o}^\infty \int_0^{2\pi}
           {Q(r') \over \pi} \, {z^{2.7} \over
            [(r^2+r'^2+z^2-2rr'\cos\phi)^{1/2}]^{4.7} }
           \, r' d\phi \, dr'
\end{equation}
and
\begin{equation}
\label{equ_sr}
S_r(r,z) = \int_{r_o}^\infty \int_0^{2\pi}
           {Q(r') \over \pi} \, {z^{1.7}(r-r'\cos\phi) \over
            [(r^2+r'^2+z^2-2rr'\cos\phi)^{1/2}]^{4.7} }
           \, r' d\phi \, dr'
\end{equation}
where $S_z$ and $S_r$ are the corresponding components of the vector
$\vec S$ defined in equation~(\ref{equ_svec})
(the $S_\phi$ component is zero due to the axial symmetry of the
accretion disk)
and $Q(r')$ is the radiation emission per area of the disk calculated
through equation~(\ref{equ_qdisk}).

Thus we have 
\begin{equation}
\vec{f}_{rad} = {\sigma \over c} \,
                k(r,z,t) \left( { 1 \over \rho \sigma v_{th}}
                \left| {dv_z \over dz} \right| \right)^{\alpha (r,z,t)}
                [ S_z(r,z) \hat z + S_r(r,z) \hat r]
\end{equation}
where $S_z$ and $S_r$ are given by equations~(\ref{equ_sz}) and
(\ref{equ_sr}).

We also account for the fact that,
as discussed by \citet{cas75} for OB star winds,
if the velocity gradient increases to sufficiently high values,
or the density of the wind decreases to sufficiently low values,
the contribution for the radiation pressure due to lines arrives at a
maximum value.
Similar to the case of OB stars \citep{abb82},
we found that the maximum value of the line radiation
pressure was obtained when
$(1/\rho \sigma v_{th}) \, dv_z/dz \lesssim 10^7$.
In this work we have assumed that the maximum possible value of the
line radiation pressure is obtained when
$(1/\rho \sigma v_{th}) \, dv_z/dz = 10^7$
(rather than $10^8$ as we did in Paper~2).

Therefore the expression for the total line radiation force per mass
for disk winds which we adopt is
\begin{equation}
\vec{f}_{rad} = {\sigma \over c} \,
              k(r,z,t) \left( \max \left[{ 1 \over \rho \sigma v_{th}}
              \left| {dv_z \over dz} \right|\, ,\, 10^7 \right] \,
              \right)^{\alpha(r,z,t)}
              [ S_z(r,z) \hat z + S_r(r,z) \hat r] \; .
\end{equation}

In order to calculate the $k(r,z,t)$ and $\alpha(r,z,t)$ functions
self consistently throughout the hydrodynamic calculations described
below and simultaneously keep the computation manageable the
numerical code was written such that the spatial distribution of
the $k$ and $\alpha$ are calculated for values of number density
varying from $10^6 \, {\rm cm}^{-3}$ to
$10^{10} \, {\rm cm}^{-3}$ at each computational spatial grid point.
A fiducial temperature is taken for these calculations,
since as we find in the models presented here,
in inner disk region where the line radiation force is most
crucial,
the temperature does not change significantly
due to radiative heating and cooling.
During the hydrodynamic calculations,
the actual values of $k$ and $\alpha$ are obtained through
geometric interpolation from the above values and the
current value of density at each computational grid point.

For the continuum radiation pressure the total radiation flux is
calculated throughout the disk wind:
\begin{equation}
\vec F = \oint \left[ \, {\int_0^\infty I(\hat{n}) d\nu} \,
                \right] \hat{n} \, d\Omega \; .
\end{equation}
Taking the source of radiation to be the accretion disk, we have
\begin{equation}
\label{equ_cfz}
F_z(r,z) = \int_{r_o}^\infty \int_0^{2\pi}
           {Q(r') \over \pi} \, {z^2 \over
            [(r^2+r'^2+z^2-2rr'\cos\phi)^{1/2}]^4 }
           \, r' d\phi \, dr'
\end{equation}
and
\begin{equation}
\label{equ_cfr}
F_r(r,z) = \int_{r_o}^\infty \int_0^{2\pi}
           {Q(r') \over \pi} \, {z (r-r'\cos\phi) \over
            [(r^2+r'^2+z^2-2rr'\cos\phi)^{1/2}]^4 }
           \, r' d\phi \, dr' 
\end{equation}
where $F_z$ and $F_r$ are the corresponding components of the radiation
flux vector $\vec F$ originating from the disk (the $F_\phi$ component
is zero due to the axial symmetry of the accretion disk) and $Q(r')$ is
the radiation emission per area of the disk calculated through
equation~(\ref{equ_qdisk}).

\section{Radiative Heating and Cooling}
\label{sec_rhc}

Radiative heating and cooling are computed in our model assuming an
optically thin disk spectrum.
A local balance between heating and cooling is assumed,
subject to the constraint of particle number conservation for
each element.
The ionization processes included in the calculation are
photoionization,
charge transfer ionization,
and collisional ionization.
Recombination processes include radiative and dielectronic
recombination and charge transfer.
The heating terms includes photoionization heating,
Compton heating, charge transfer, and collisional de-excitation.
Cooling terms include radiative and dielectronic recombination,
bremsstrahlung,
collisional ionization,
collision excitation of bound levels,
and (endothermic) charge transfer.
Details of the radiative heating and cooling calculations are
described in detail by \citet{kal82}.

\begin{figure}
\epsscale{1.0}
\plotone{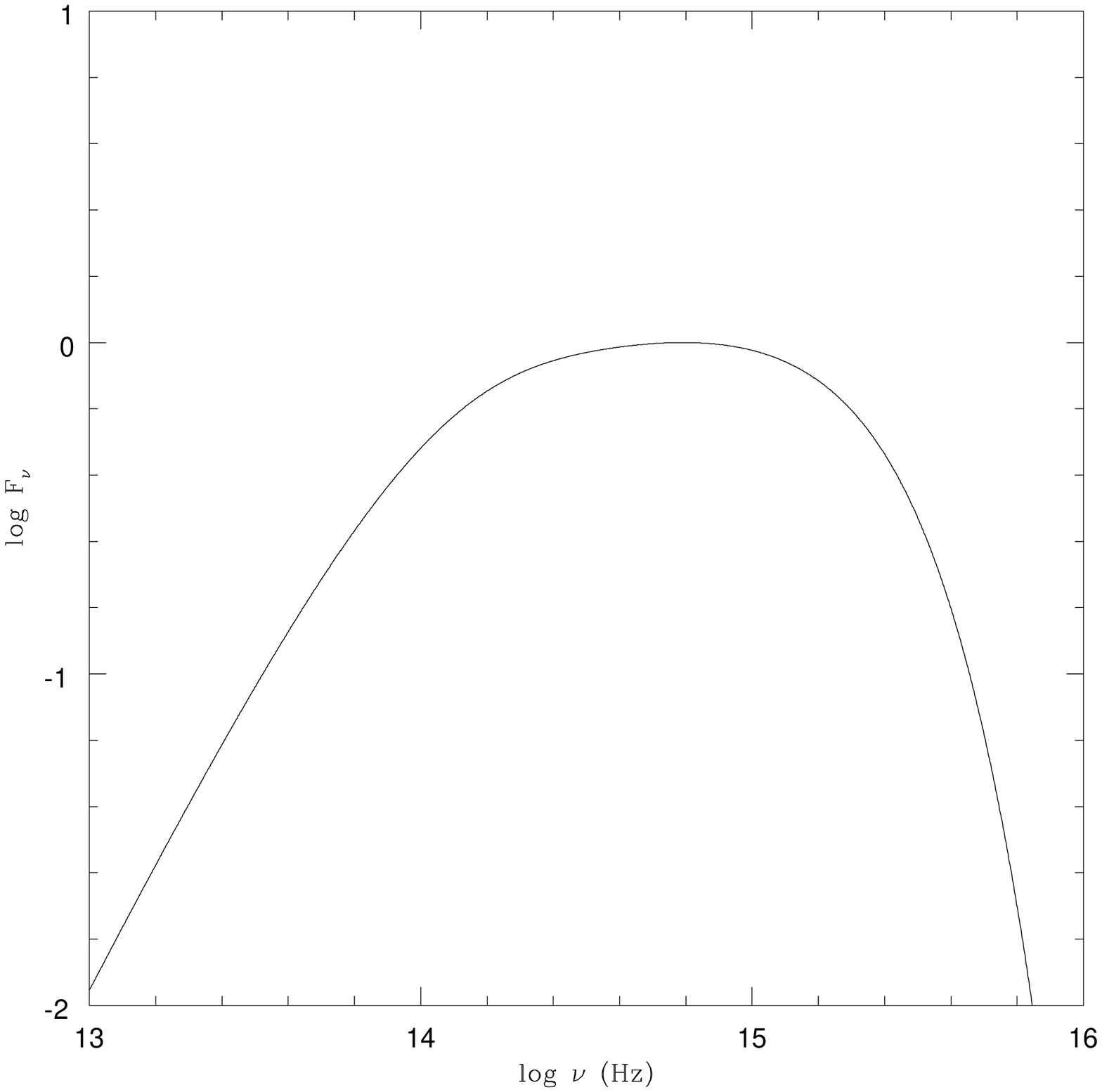}
\caption{
Logarithm of radiation flux vs logarithm of
frequency in units of ${\rm Hz}$ for the accretion disk used in this
work.
This is the spectrum distribution used for the radiative heating
and cooling calculations.
The physical parameters used here are $L_{disk}=L_{\sun}$ and
$R_{wd} = 0.01 R_{\sun}$.
}
\label{fig_spe}
\end{figure}

Disk models have succeeded in accounting for the observed
continuum spectrum in many CVs
\citep[e.g.][]{her79,kip79,kip80,bat80,may80,kra81,has85,wad88,lad89,
               dia96,wad98}.
The form of the spectrum applied in this work for the radiative heating
and cooling calculations is obtained by integrating contributions
over the disk surface calculated from equation~(\ref{equ_qdisk}),
assuming local blackbody emission,
and viewing the disk at infinity.
In Figure~\ref{fig_spe} we show the disk spectrum used for the
radiative heating and cooling calculations.
We note that an assumption made in the radiative cooling
and heating calculations is that the form of the spectrum
is position independent,
and thus the rates calculated here will be more accurate
higher above the disk.
In the future we will introduce more accurate heating and cooling
rates into our models.

\begin{figure}
\epsscale{1.0}
\plotone{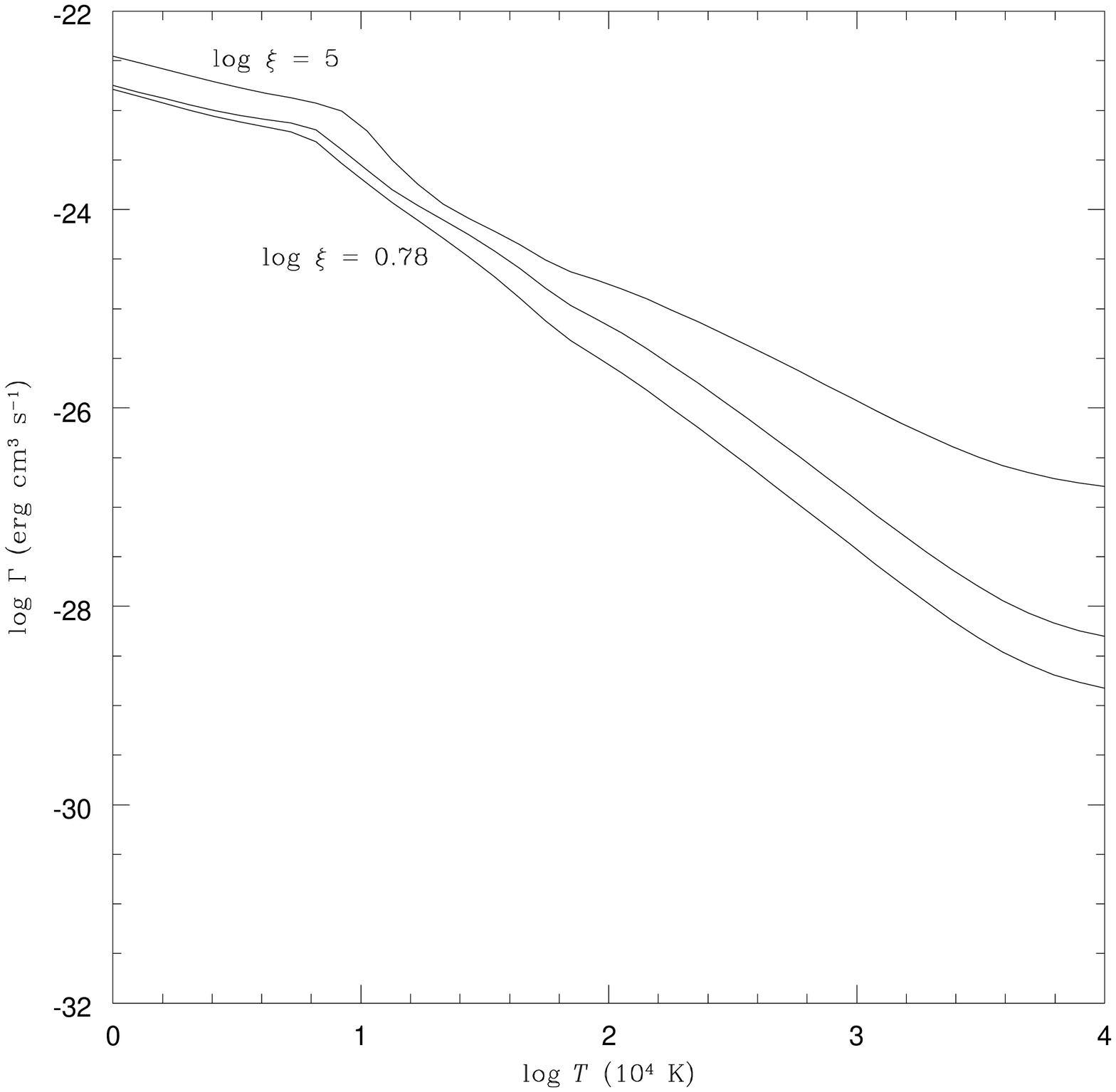}
\caption{
Logarithm of the radiative heating rate in units of
${\rm erg} \, {\rm cm}^3 \, {\rm s}^{-1}$ vs the logarithm of
the temperature in units of $10^4 \; {\rm K}$.
The ionization parameters used are log $\xi$ = 5.0, 2.3, 0.78
}
\label{fig_hea}
\end{figure}

\begin{figure}
\epsscale{1.0}
\plotone{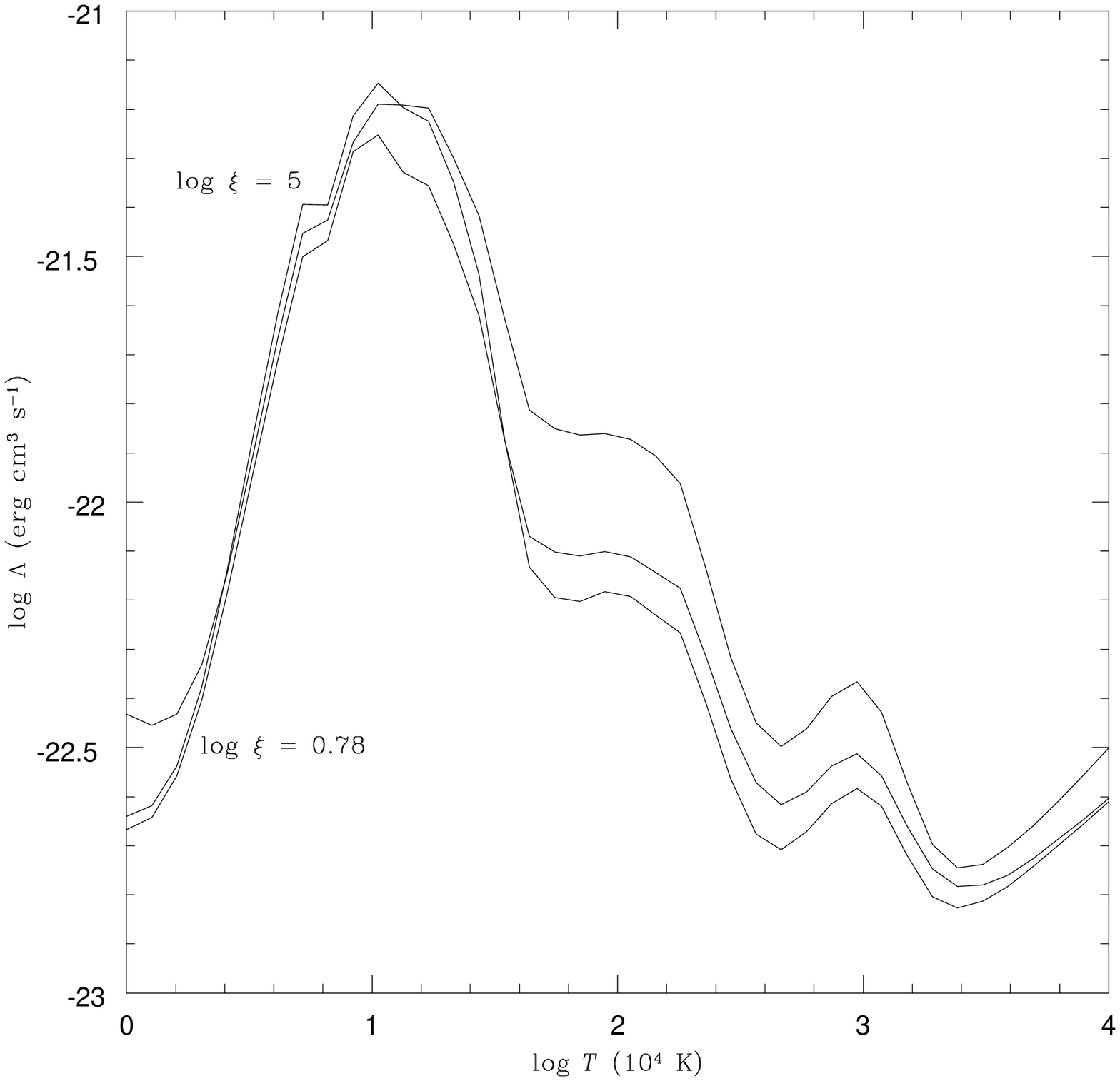}
\caption{
Logarithm of the radiative cooling rate in units of
${\rm erg} \, {\rm cm}^3 \, {\rm s}^{-1}$ vs the logarithm of
the temperature in units of $10^4 \; {\rm K}$.
The ionization parameters used are log $\xi$ = 5.0, 2.3, 0.78
}
\label{fig_coo}
\end{figure}

In Figure~\ref{fig_hea} and Figure~\ref{fig_coo} the radiative heating
and cooling are shown respectively as a function of temperature for
different values of the ionization parameter $\xi$ defined as:
\begin{equation}
\xi \equiv {4 \pi | \vec{F} | \over n}
\end{equation}
where $F$ is the radiation flux and $n$ the number density.

From Paper~2 we found that the wind temperature of CV winds
range in values of up to $\sim 40,000 \, {\rm K}$.
Thus in our work here we have approximated the heating and cooling
rates the results presented in Figure~\ref{fig_hea} and
Figure~\ref{fig_coo} by analytical expressions that agree with the
detailed calculation within 20\% for the heating rates and within
30\% for the cooling rates for temperatures of up to
$50,000 \, {\rm K}$.

For the radiative heating rate we implemented the following expression
(in units of ${\rm erg} \, {\rm cm}^3 \, {\rm s}^{-1}$):
\begin{equation}
\Gamma = 1.6 \times 10^{-21} \; T^{-1/2}
         \left( 1 - {T \over 1.3 \times 10^5} \right)
\end{equation}
where $T$ is the absolute temperature,
valid for temperatures of up to $50,000 \, {\rm K}$.

For the radiative cooling rate we implemented the following expression
(in units of ${\rm erg} \, {\rm cm}^3 \, {\rm s}^{-1}$):
\begin{equation}
\Lambda = 1.9 \times 10^{-29} \; T^{3/2}
\end{equation}
where $T$ is the absolute temperature,
valid for temperatures of up to $50,000 \, {\rm K}$.

\section{Hydrodynamic Model}
\label{sec_hyd}

Our hydrodynamic model uses the Piece-Wise Parabolic Method (PPM)
numerical scheme \citep{col84}.
We use the 2.5-dimensional hydrodynamic equations of the wind in
cylindrical coordinates;
i.e.,
the three-dimensional hydrodynamic equations are reduced to
two-dimensional equations by assuming that the derivative of any
physical variable with respect to $\phi$ is zero.
The origin of the coordinate system is located at the center of
the accretion disk and the ``z=0'' plane is located to coincide
with the accretion disk midplane.
The hydrodynamic equations are presented in the Appendix.

In this treatment we have implemented the energy equation self
consistently.
In particular,
the adiabatic heating and cooling effects are implemented through the
first and second term on the right hand side of
equation~(\ref{equ_hydro_e}),
and the radiative heating and cooling rates are implemented through
the last term on the right hand side of equation~(\ref{equ_hydro_e}).

As discussed in section~\ref{sec_rf},
our treatment of the line radiation pressure assumes negligible the
$\phi$ component of the line radiation force,
and therefore the specific angular momentum is conserved within
our models.
We note that strictly,
the azimuthal component of the line radiation force is not required
to be zero,
and thus the angular momentum is not necessarily conserved,
but this is likely a small correction.

As we discovered in Paper~1 and Paper~2,
care must be taken with the boundary conditions to avoid
numerical instabilities.
In this work we treat the boundary conditions in a form
equivalent to that discussed in Paper~2.
Our spatial grids are also equivalent to those explored and
discussed in Paper~2.

In the models presented here we solve the hydrodynamic equations
over a range of radii from $0.02 R_{\sun}$
($2R_{wd}$ in our models) to $1.5 R_{\sun}$.
From equation~(\ref{equ_qdisk}) it is found that the energy emission
per area $Q$ will increase slightly as radius decreases from
$2R_{wd}$ down to $\approx 1.36 R_{wd}$  and will decrease
strongly as radius decreases from $\approx 1.36 R_{wd}$
down to the white dwarf radii ($R_{wd}$).
In Paper~1 we showed that the wind density tends to decrease
with decreasing radiation flux.
Thus the mass loss rate for radii less that $2 R_{wd}$ will not
alter significantly the total mass loss rate of the wind.
Also,
in Paper~2,
we presented models with spatial grids with ``r'' down to $R_{wd}$
and found essentially the same results with models with ``r'' down
to $2R_{wd}$ with model parameters otherwise identical.

\section{Results}
\label{sec_res}

In Table~\ref{tab} we present the wind mass loss rates from our
previous models and for the models presented here.
The second column indicates the dimension of the model,
2.5D are the full 3D hydrodynamic equations under
azimuthal symmetry;
i.e. the derivatives with respect to the azimuthal angle are
assumed zero. 
2.5D models allow a consistent treatment of the 3D equations
(under azimuthal symmetry) while mathematically having only two
independent spatial directions,
keeping the calculations more manageable.
The third column indicates the radial disk structure assumed.
In all models a steady disk is assumed.
An ``I'' indicates an isothermal disk (no radial dependence).
An ``S'' indicates the radial dependence of disk radiation emission
and temperature of a standard Shakura-Sunyaev disk.
The third and fourth columns indicate the values of the $k$ and
$\alpha$ parameters used in each model.
In our previous models (A-F) we had assumed constant ionization
equilibrium throughout the wind,
and therefore constant values for these two parameters.
In our current work,
as described above,
we implement the spatial dependence (Model G) and
the spatial and time dependence (Model H) of these two parameters
and find values that vary throughout our spatial grids 
(and in time for Model H) within the intervals indicated in
Table~\ref{tab}.

\begin{deluxetable}{lcccccrc}
\tabletypesize{\scriptsize}
\tablecolumns{8}
\tablecaption{CV Accretion Disk Wind Mass-Loss Rates for Different
              Models}
\tablehead{
  \colhead{Model} &
  \colhead{Dim} &
  \colhead{Disk} &
  \colhead{$k$} &
  \colhead{$\alpha$} &
  \colhead{Comments} &
  \colhead{$\dot M$($M_{\sun} \; yr^{-1}$)} &
  \colhead{Reference} }
\startdata
A & 1 & I & 1/30 & 0.7 &
Isothermal Wind &
$2\times 10^{-14}$ &
Paper~1 \\
B & 2.5 & I & 1/30 & 0.7 &
Model A extended to 2.5D; &
$2\times 10^{-14}$ &
Paper~1 \\
  &  &  &  &  &
radiation pressure in ``r'' neglected & & \\
C & 2.5 & I & 1/3 & 0.7 &
Model B with an increased value of $k$ &
$6\times 10^{-14}$ &
Paper~2 \\
D & 2.5 & I & 1/3 & 0.7 &
Model C with a maximum cutoff &
$2\times 10^{-13}$ &
Paper~2 \\
  &  &  &  &  &
value for the line pressure implemented & & \\
E & 2.5 & I & 1/3 & 0.7 &
Model D with an adiabatic wind &
$2\times 10^{-13}$ &
Paper~2 \\
  &  &  &  &  &
(rather than an isothermal wind) & & \\
F & 2.5 & S & 1/3 & 0.7 &
Model E with a &
$8\times 10^{-12}$ &
Paper~2 \\
  &  &  &  &  &
standard Shakura-Sunyaev disk & & \\
  &  &  &  &  &
(rather than isothermal disk) & & \\
  &  &  &  &  &
radiation pressure in ``r'' included & & \\
G & 2.5 & S & 0.03-0.5 & 0.6-0.8 &
Model F with spatial &
$2\times 10^{-11}$ &
current work \\
  &  &  &  &  &
dependence of $k$ and $\alpha$ derived & & \\ 
  &  &  &  &  &
from the spatial distribution & & \\
  &  &  &  &  &
of radiation field but assuming & & \\
  &  &  &  &  &
a fiducial number density of $10^8 \; {\rm cm^{-3}}$ & & \\
H & 2.5 & S & 0.03-0.6 & 0.55-0.85 &
Model F with spatial and time &
$4\times 10^{-12}$ &
current work \\
  &  &  &  &  &
dependence of $k$ and $\alpha$ derived from & & \\
  &  &  &  &  &
the spatial distribution of radiation and the & & \\
  &  &  &  &  &
spatial and time distribution of mass density & & \\
\enddata
\label{tab}
\end{deluxetable}

In implementing local ionization equilibrium in our models,
we first run models that include the spatial dependence of the
radiation field above the disk but assume a fiducial number density
of $10^8 \;{\rm cm}^{-3}$ (Model~G).
From this we obtain the values of the ``$k$'' and ``$\alpha$'' line
radiation parameters for each point of the computational grid.
Results from these runs are presented in Figure~\ref{fig_mg}.
The values of $k$ we found varied from $0.03$ to $0.5$ and
values of $\alpha$ varying from $0.6$ to $0.8$.
The higher values of both $k$ and $\alpha$ are obtained in the
lower inner region of the wind (lower left hand side of
Figure~\ref{fig_mg}).
The lower values of both $k$ and $\alpha$ are obtained in the
lower outer region of the wind (lower right hand side of
Figure~\ref{fig_mg}).
At $z=R_{\sun}$,
the $k$ parameter was found to vary from $0.16$ to $0.26$ and
the $\alpha$ parameter was found to vary from $0.66$ to $0.70$.
In this model we find similar terminal velocities with respect
to the models presented in Paper~2,
but densities values increased by a factor of 2.
We find steady wind flows coming from the disk.
Wind-mass loss rates calculated in this work are obtained by
integrating the mass flux at the start of the wind,
at the lower boundary of the spatial computational grid.
The total wind mass loss rate we find for this model is
$2\times 10^{-11} \; M_{\sun} \; {\rm yr}^{-1}$.
This increase in the wind mass loss rate is due to the increase
of the $k$ in the lower inner region where the radiation flux
is stronger.

In order to calculate the $k(r,z,t)$ and $\alpha(r,z,t)$ functions
self consistently throughout the hydrodynamic calculations,
as we described in \S\ref{sec_rhc},
the spatial distributions of the $k$ and $\alpha$ are calculated
in a similar manner as in the model presented in Figure~\ref{fig_mg},
but now for values of number density varying from
$10^6 \; {\rm cm}^{-3}$ to $10^{10} \; {\rm cm}^{-3}$
at each computational spatial grid point (Model H).
A fiducial electron temperature of $30,000 \, {\rm K}$ is taken for
these calculations.
During the hydrodynamic calculations,
the actual values of $k$ and $\alpha$ are obtained through
geometric interpolation from the above values and the
current value of density at each computational spatial grid point.
Results from these models are presented in Figure~\ref{fig_mh}.
For the above range of densities at each spatial grid point,
we find values of $k$ varying from $0.03$ to $0.6$ and
values of $\alpha$ varying from $0.55$ to $0.85$.
Again we find steady wind flows coming from the disk.
In the lower inner region,
due to the increase of density the $k$ parameter drops by
a factor of $\sim 2$,
producing in turn a decrease of total wind mass rate.
The wind mass loss rate for this model is
$4\times 10^{-12} \, M_{\sun} \, {\rm yr}^{-1}$.
The velocity fields are very similar to the previous models,
finding terminal velocities in the order of
$\sim 3000 {\rm \, km \, s}^{-1}$.
We note that the mass loss rate is a approximately a factor of two
below the values we found in our earlier models in Paper~2.
This difference is due to the detailed calculations of the
line radiation force parameters,
which when the local density values at the
inner region are considered,
we find slightly lower values of the $k$ line radiation force
parameter.

In Figures~\ref{fig_mkh} and \ref{fig_mah} we present contour plots of
the line radiation force parameter $k$ and $\alpha$ respectively,
once the calculations arrive at a steady solution (Model~H).
These two figures illustrate the coupling between line force
parameters,
ionization equilibrium,
and wind dynamics within an accretion disk wind.
We note that,
although the ionization structures are coupled with the hydrodynamics,
within typical CV parameters
($M_{wd} = 0.6 M_{\sun}$;
 $R_{wd}=0.01R_{\sun}$;
 $L_{disk} = L_{\sun}$),
in the higher wind density areas,
the line force parameters do not change significantly;
thus justifying the assumption of our earlier models of constant
ionization equilibrium
(and thus constant line force parameters).
We also note that in QSOs/AGN,
the presence of strong x-ray emission will generate strong
changes in the ionization balance of the wind,
and the assumption of constant ionization equilibrium in the wind
will no longer hold \citep{mur95,hil02}.

Studies of observed ultraviolet line profiles from CVs have estimated
CV  wind mass loss rates to be in the order of
$10^{-12}$ to $10^{-11} M_{\sun} \, {\rm yr}^{-1}$ and found CV wind
terminal speeds in the order of $\sim 3000 \, {\rm km} \, {\rm s}^{-1}$
\citep[e.g.][]{kra81,cor82,gre82,mau87,pri95,fri97}
thus the models developed in this work finds values for
wind mass loss rates and terminal speeds consistent with observations.

In the models presented in this work we also implement radiative
heating and cooling. In Figure~\ref{fig_mth} we present a velocity
field graph superimposed with temperature contours for the local
ionization equilibrium disk wind model.
Comparing with the results from Paper~2 (Figure~\ref{fig_mtf}),
we find that radiative heating and cooling tends to maintain
the wind temperature in the higher density regions near
the disk,
rather than cooling due to adiabatic expansion as we had found
in our earlier models of Paper~2.
Therefore the radiative heating and cooling processes are playing
an important role in the temperature structure of the wind.
We are finding higher overall temperatures than in Paper~2
where we neglected radiative heating and cooling processes
and assumed only adiabatic heating cooling processes.

As a consistency check,
in Figure~\ref{fig_mc1} we present the results of models with a
computational spatial grid that starts at $r=R_{wd}$
(rather than $r=2.0R_{wd}$ as in the other models discussed previously
in this work).
In Figure~\ref{fig_mc2} we present results from models with a
computational spatial grid starting at $r=R_{wd}$ and with an initial
velocity at the base of the wind of $1 \, {\rm km \, s}^{-1}$
(rather than $10 \, {\rm km \, s}^{-1}$ as in the other models
discussed previously in this work).
In both cases we find similar overall results:
a steady biconical wind velocity/density structure,
with similar velocity and density values.
We note that these results although similar,
are not identical to those of Figure~\ref{fig_mh},
and thus represent numerical effects since the physical
parameters have remained identical.
However the deviations from Figure~\ref{fig_mh} are considerably
smaller than the changes of results we have found in earlier work
when we physical parameters have been modified as our
models have developed (Paper~2),
thus we are confident in our overall results.
In future work we shall study these numerical effects in greater
detail;
a precise knowledge of these model limitations may be important
to achieve precise comparisons with observations.

As we had found in our earlier models of Paper~1 and Paper~2,
rotational forces are important in the study of winds from
accretion disks.
They cause the velocity streamlines to collide which results in
an enhanced density region.
In this region the wind speed is reduced and the wind density is
increased.
The increase in density caused by the collision of streamlines is
important because it permits the appearance of blue-shifted absorption
lines as observed in P~Cygni profiles of low-inclination CVs.

\section{Summary and Conclusions}
\label{sec_sum}

We have developed a 2.5-dimensional hydrodynamic line-driven
accretion disk wind model.
Our model solves a complete set of adiabatic hydrodynamic partial
differential equations,
using the PPM numerical scheme and implementing the radial temperature
and radiation emission distributions on the surface of an accretion
disk.
Our models calculate the line radiation force parameters throughout
the wind at each computational grid point for each time step
through the calculation of line opacities from atomic data,
similar to the calculations presented by \citet{abb82} for OB stars.
We treat the hydrodynamic equation of energy self consistently
including radiative heating and cooling as well as adiabatic
expansion and compression.

We find steady wind flows coming from the disk
with similar density and velocity structures as our earlier models
of Paper~2 in which we had not yet included radiative heating and
cooling and in which we had assumed constant ionization equilibrium
(therefore constant values for the line radiation force parameters).
The steady nature of our solution contrasts with the strong
fluctuations in density and velocity found by \citet{pro98}
for CV winds for physical parameters similar to the ones
we used here.
A possibility that could account for the difference in results from
our models and those of \citet{pro98}
is the treatment of the lower boundary conditions.
In future work we shall explore this possibility in detail.

We note that as our numerical models evolve towards more
realistic ones,
it may result that a more accurate treatment of the line radiation
force (which would include line-overlapping, rather than
assuming the Sobolev approximation for all lines)
or a more accurate model of the accretion disk
(which may result in an unsteady picture of the accretion disk
rather than the standard Shakura-Sunyaev disk)
may lead to physical instabilities.
In our systematic approach to the accretion disk wind problem,
we started with an analytic 1D model,
which although being our ``crudest'' one,
did allow for an analytic solution which has proven to be
invaluable as a test case for our numerical algorithms.
From this initial model we have included,
on a step by step basis (Paper~1, Pereyra~1997, Paper~2)
elements such as a more accurate treatment of the
line radiation force (in which we have always applied the Sobolev
approximation),
radial structure of the accretion disk,
adiabatic heating and cooling processes, etc.
This approach has allowed us to establish clearly a relationship
between the elements we have introduced (both of numerical and
physical nature) and our results,
as our models have developed into more realistic ones.
In this sense we are confident that if intrinsic physical instabilities
should appear as we continue our work,
we will be able to establish their physical origin.

From our models we calculate wind mass-loss rates
and terminal velocities.
For typical cataclysmic variable parameter values of disk luminosity
$L_{disk}=L_{\sun}$,
white dwarf mass $M_{wd}=0.6M_{\sun}$,
and white dwarf radii $R_{wd}=0.01R_{\sun}$,
we obtain a wind mass-loss rate of
$\dot M_{wind}=4 \times 10^{-12} M_{\sun} \, {\rm yr}^{-1}$,
and a terminal velocity of $\sim 3000 {\rm \, km \, s}^{-1}$.
We note that the mass loss rate is a approximately a factor of two
below the values we found in our earlier models in Paper~2.
This difference is due to the detailed calculations of the
line radiation force parameters,
which when the local density values at the
inner region are considered,
we find slightly lower values of the $k$ line radiation force
parameter.

The results in this work suggest that the details of the ionization
equilibrium in CV disk winds,
within reasonable parameters,
may not have considerable effect over the velocity fields of the wind
nor in the wind mass distribution.
Although these details have to be taken into account in order to derive
an accurate theoretical value of the wind mass loss rate as
illustrated by the difference by a factor of two between the models
presented here and the earlier ones of Paper~2.
The similar wind mass distribution and velocity fields found here
with respect to those of Paper~2,
would indicate a similar inclination angle dependence of the
\ion{C}{4} 1550 \AA \  line.
But the decrease in total wind mass loss rate would generate
weaker line features.
In QSOs, where the ionization balance of the gas material may 
change significantly,
the coupling between the ion populations and the radiative
hydrodynamics will have a important effects in the overall
wind dynamics;
thus a model capable of representing this coupling becomes necessary.

A shortcoming of our models is that,
at each spatial point for each time step
we assume direction-independent values for the line
radiation parameters
and we also assume position independent values for the
radiative heating and cooling parameters.
In future work we plan to include these effects.
We also plan to extend our disk wind models to low mass X-ray binaries
and QSOs/AGN,
where the local ionization equilibrium
will play a more crucial role in the overall dynamics.

\acknowledgments

We thank John Hillier and Dave Turnshek for their useful comments and 
discussions.
We also thank the referee for useful comments and suggestions that
have significantly improved the presentation of this work.
This work was supported by he National Science Foundation
under Grant AST-0071193.

\appendix

\section{Hydrodynamic Equations}

The hydrodynamic equations implemented in the models presented are,
the equation of state:
\begin{equation}
P = (\gamma -1 )\rho e
\end{equation}                  
the mass conservation equation:
\begin{equation}
{\partial\rho \over \partial t}
  + {1 \over r}{\partial(r \rho v_r) \over \partial r}
  + {\partial(\rho v_z) \over \partial z} = 0
\end{equation}
the momentum conservation equations:
\begin{eqnarray}
\rho \, {\partial v_r \over \partial t}
  + \rho \, v_r{\partial v_r \over \partial r}
& - & \rho \, {{v_\phi}^2 \over r}
  + \rho \, v_z{\partial v_r \over \partial z}
  = \\
&   & \nonumber \\ 
& - & \rho \, {GM_{wd} \over (r^2 + z^2)} \, {r \over (r^2+ z^2)^{1/2}}
  - {\partial P \over \partial r} + \rho \, {\sigma F_r(r,z) \over c}
    \nonumber \\
&   & \nonumber \\
& + &  \rho \, {\sigma S_r(r,z) \over c} \,
         k(r,z,t) \left( \max \left[{ 1 \over \rho \sigma v_{th}}
         \left| {dv_z \over dz} \right|,10^7 \right] \;
         \right)^{\alpha(r,z,t)}  \nonumber \\
&   & \nonumber
\end{eqnarray}
\begin{equation}
\rho \, {\partial v_\phi \over \partial t}
  + \rho \, v_r{\partial v_\phi \over \partial r}
  + \rho \, {v_\phi v_r \over r}
  + \rho \, v_z{\partial v_\phi \over \partial z}
  = 0
\end{equation}
\begin{eqnarray}
&   & \nonumber \\
\rho \, {\partial v_z \over \partial t}
  + \rho  \, v_r{\partial v_z \over \partial r}
& + & \rho \, v_z{\partial v_z \over \partial z}
 = \\
&   & \nonumber \\
& - & \rho \, {GM_{wd} \over (r^2+z^2)} \, {z \over (r^2+z^2)^{1/2}} 
 - {\partial P \over \partial z} + \rho \, {\sigma F_z(r,z) \over c} 
   \nonumber \\
&   & \nonumber \\
& + & \rho \, {\sigma S_z(r,z) \over c} \,
         k(r,z,t) \left( \max \left[{ 1 \over \rho \sigma v_{th}}
         \left| {dv_z \over dz} \right|, 10^7 \right] \;
         \right)^{\alpha(r,z,t)} \nonumber
\end{eqnarray}
and the energy conservation equation:
\begin{eqnarray}
\label{equ_hydro_e}
{\partial \rho E \over \partial t}
+ {1 \over r} \, {\partial r \rho E v_r \over \partial r}
& + & {\partial \rho E v_z \over \partial z}
  = \\
&   & \nonumber \\
& - & {1 \over r} \, {\partial r P v_r \over \partial r}
  -{\partial P v_z \over \partial z} \nonumber \\
&   &  \nonumber \\
& - & \rho \, {GM_{wd} \over (r^2 + z^2)} \,
              {r \over (r^2+ z^2)^{1/2}} \, v_r 
 + \rho \,  {\sigma F_r(r,z) \over c} \, v_r \nonumber \\
&   &  \nonumber \\
& + & \rho \, {\sigma S_r(r,z) \over c} \,
         k(r,z,t) \left( \max \left[{ 1 \over \rho \sigma v_{th}}
         \left| {dv_z \over dz} \right|,10^7 \right] \;
         \right)^{\alpha(r,z,t)} v_r \nonumber \\
&   &  \nonumber \\
& - & \rho \, {GM_{wd} \over (r^2+z^2)} \,
              {z \over (r^2+z^2)^{1/2}} \, v_z 
 + \rho \, {\sigma F_z(r,z) \over c} \, v_z  \nonumber \\
&   &  \nonumber \\
& + & \rho \, {\sigma S_z(r,z) \over c} \,
         k(r,z,t) \left( \max \left[{ 1 \over \rho \sigma v_{th}}
         \left| {dv_z \over dz} \right|, 10^7 \right] \;
         \right)^{\alpha(r,z,t)} v_z \nonumber \\
&   & \nonumber \\
& + & n^2(\Gamma(T)-\Lambda(T)) \nonumber
\end{eqnarray}
where $P$ is the pressure,
$\gamma$ is the ratio of specific heats,
$\rho$ is the density,
$e$ is the internal energy per mass,
$v_r$,
$v_\phi$,
and $v_z$ are the corresponding velocity components in cylindrical
coordinates,
$G$ is the gravitational constant,
$M_{wd}$ is the mass of the white dwarf,
$\sigma$ is the Thomson cross section per unit mass,
$F_r$ and $F_z$ are the corresponding radiation flux components
(see equations~[\ref{equ_cfz}] and [\ref{equ_cfr}]),
$c$ is the speed of light,
$v_{th}$ is the thermal velocity,
and $k$ and $\alpha$ are the line force multiplier parameters
calculated through equation~(\ref{equ_fm1}),
$S_r$ and $S_z$ are defined through
equations~(\ref{equ_sr}) and (\ref{equ_sz})
respectively,
$E = v_r^2/2 + v_\phi^2/2 + v_z^2/2 + e$ is the total energy per
mass,
$n$ is the electron number density,
$\Gamma$ is the radiative heating rate,
and $\Lambda$ is the radiative cooling rate.

\clearpage

\begin{figure}
\epsscale{0.5}
\plotone{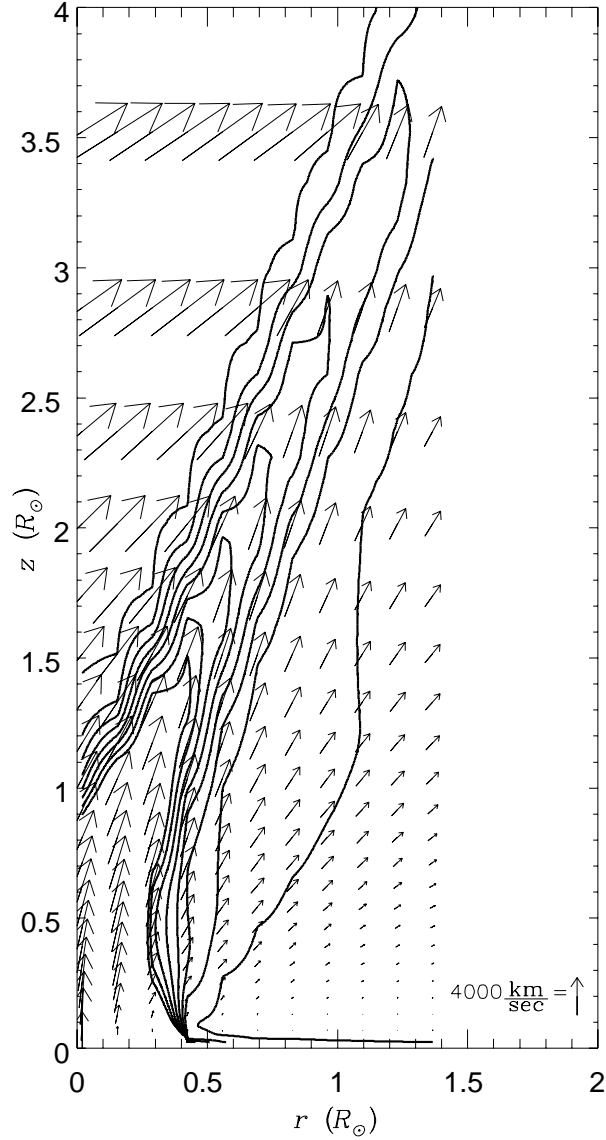}
\caption{
Vector field graph of wind velocity superimposed
with density contours for the model assuming
constant ionization equilibrium (Model~F [Paper~2]).
The line radiation force parameters applied are $k=0.3$,
$\alpha = 0.7$~.
The primary star is at the origin of the graph,
and the disk is over the horizontal axis.
The contour levels vary uniformly from a value of
$3.6 \times 10^{-16} \, {\rm g \, cm}^{-3}$ down to a value of
$0.1 \times 10^{-16} \, {\rm g \, cm}^{-3}$.
The physical parameters here used are $M_{wd} = 0.6 M_{\sun}$,
$R_{wd}=0.01R_{\sun}$,
and $L_{disk} = L_{\sun}$.
}
\label{fig_mf}
\end{figure}

\begin{figure}
\epsscale{0.5}
\plotone{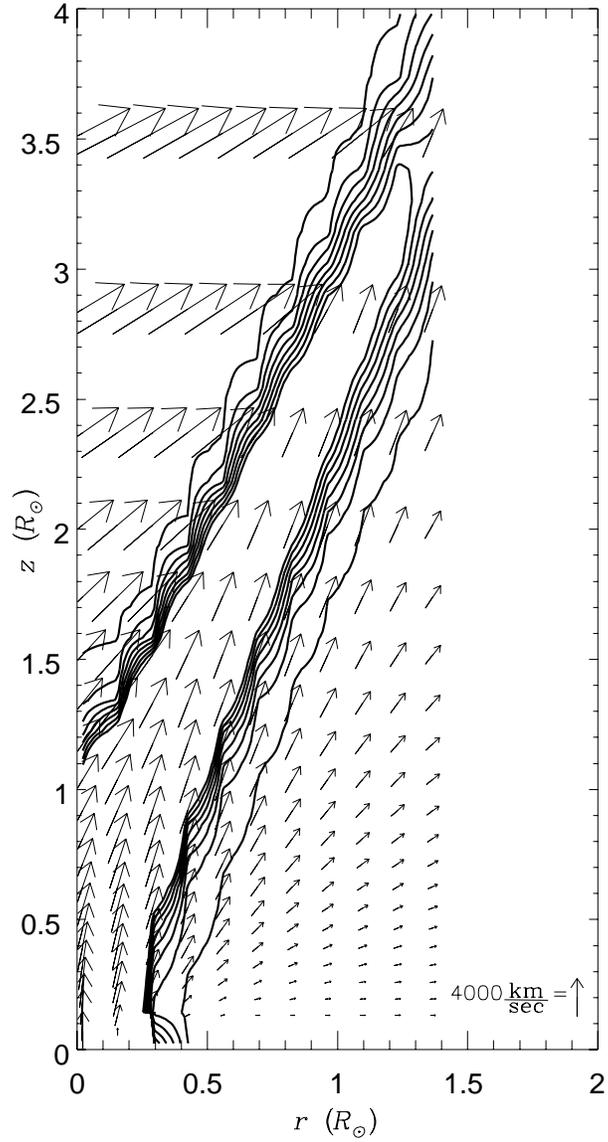}
\caption{
Vector field graph of wind velocity superimposed
with density contours for the
model assuming an ionization equilibrium
varying in space but constant in time.
A fiducial number density of $10^8 \; {\rm cm}^{-3}$ is used
for the calculations of the line radiation force parameters
(Model~G).
Parameters are similar to those of Figure~\ref{fig_mf}.
}
\label{fig_mg}
\end{figure}

\begin{figure}
\epsscale{0.5}
\plotone{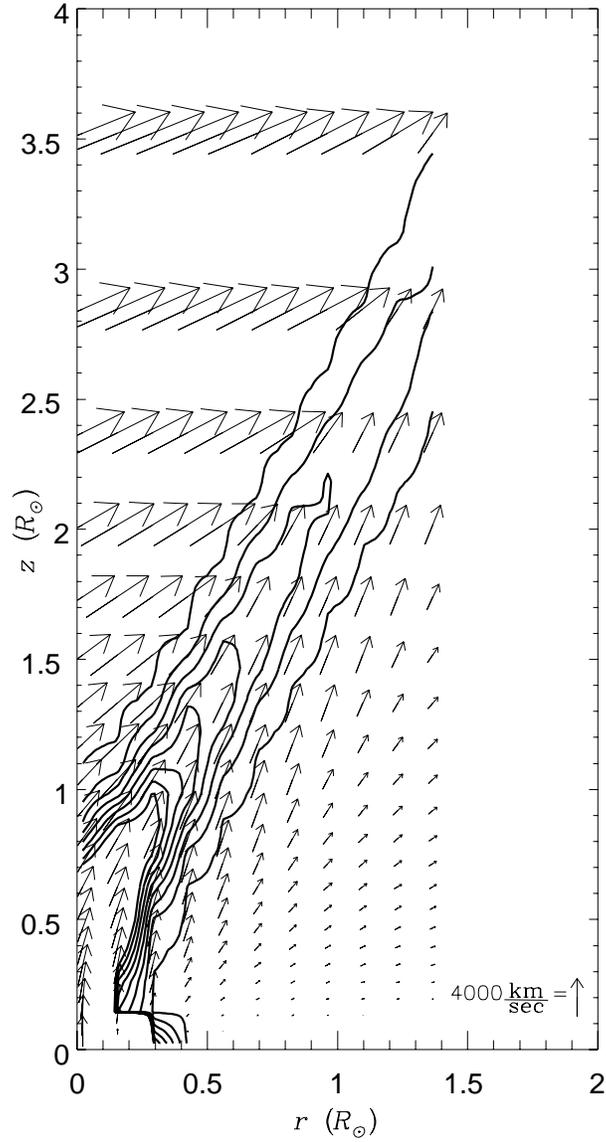}
\caption{
Vector field graph of wind velocity superimposed
with density contours for the
model applying an ionization equilibrium
varying in space and time.
The current density is applied at each spatial grid point
at each time step in the calculation of the line radiation
force parameters
(Model~H).
Parameters are similar to those of Figure~\ref{fig_mf}.
}
\label{fig_mh}
\end{figure}

\begin{figure}
\epsscale{0.5}
\plotone{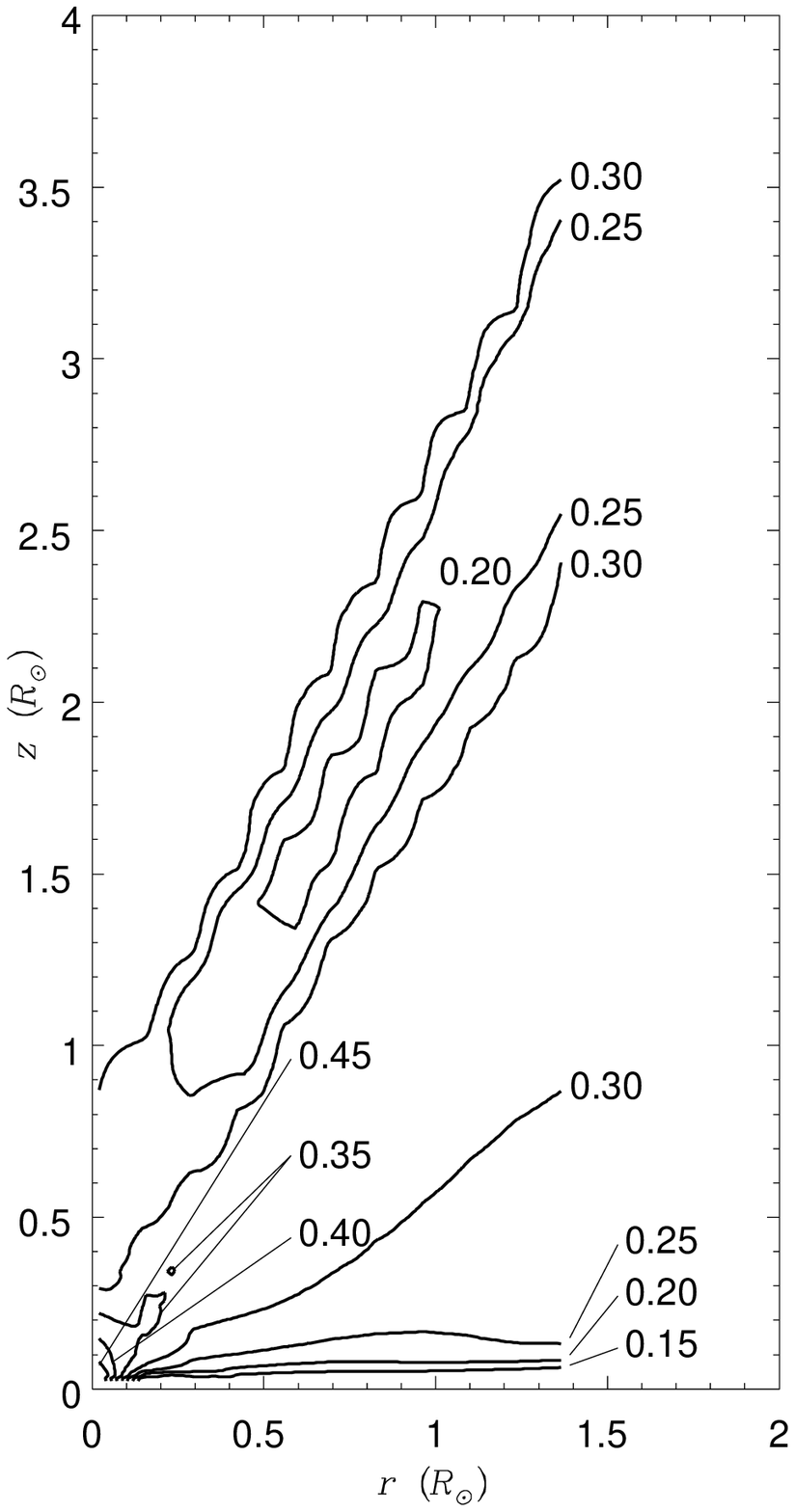}
\caption{
Contour plot of the line radiation force parameter $k$ in the
Model~H once it arrives at a steady solution.
The physical parameters used are those of Figure~\ref{fig_mf}.
}
\label{fig_mkh}
\end{figure}

\begin{figure}
\epsscale{0.5}
\plotone{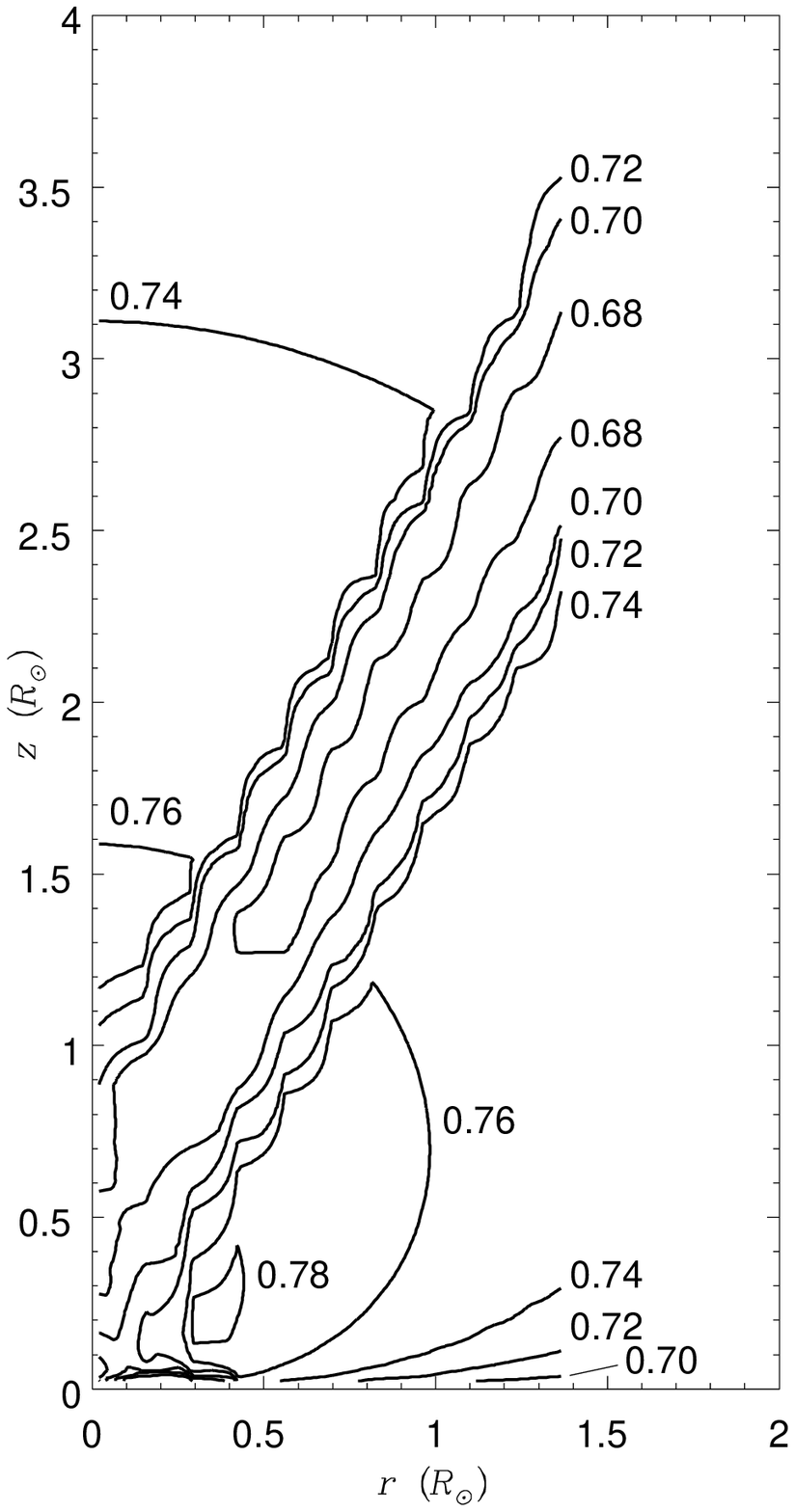}
\caption{
Contour plot of the line radiation force parameter $\alpha$ in the
Model~H once it arrives at a steady solution.
The physical parameters used are those of Figure~\ref{fig_mf}.
}
\label{fig_mah}
\end{figure}

\begin{figure}
\epsscale{0.5}
\plotone{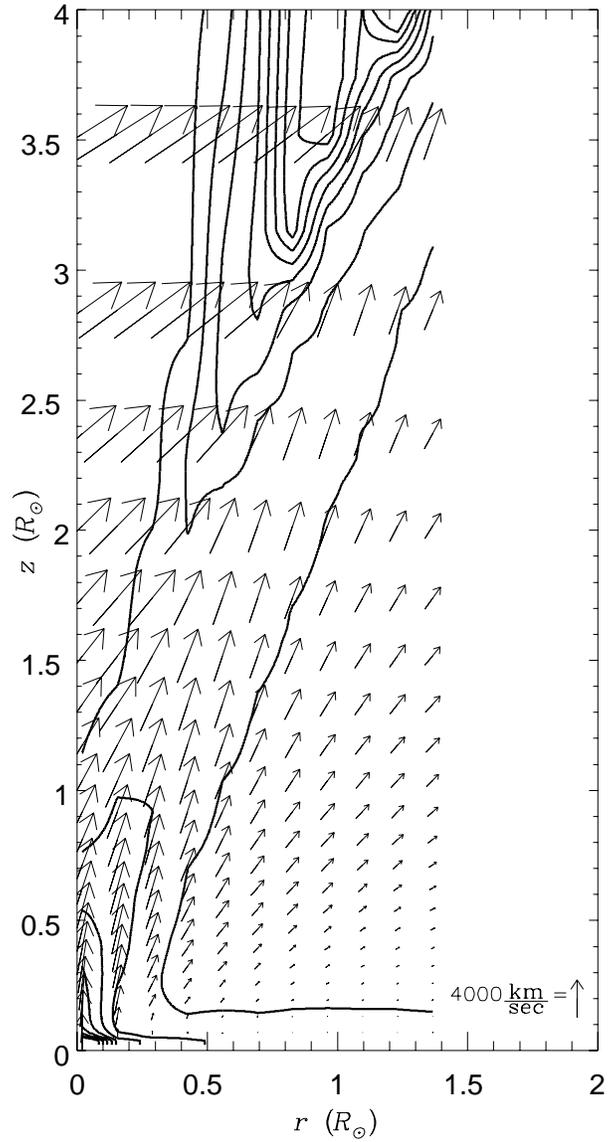}
\caption{
Vector field graph of wind velocity superimposed
with temperature contours for the
model assuming constant ionization equilibrium
and without radiative heating and cooling
(Model~F [Paper~2]).
The contour levels vary uniformly from a value of
$22,000 \, {\rm K}$  down to a value of $1,000 \, {\rm K}$.
Parameters are similar to those of Figure~\ref{fig_mf}.
}
\label{fig_mtf}
\end{figure}

\begin{figure}
\epsscale{0.5}
\plotone{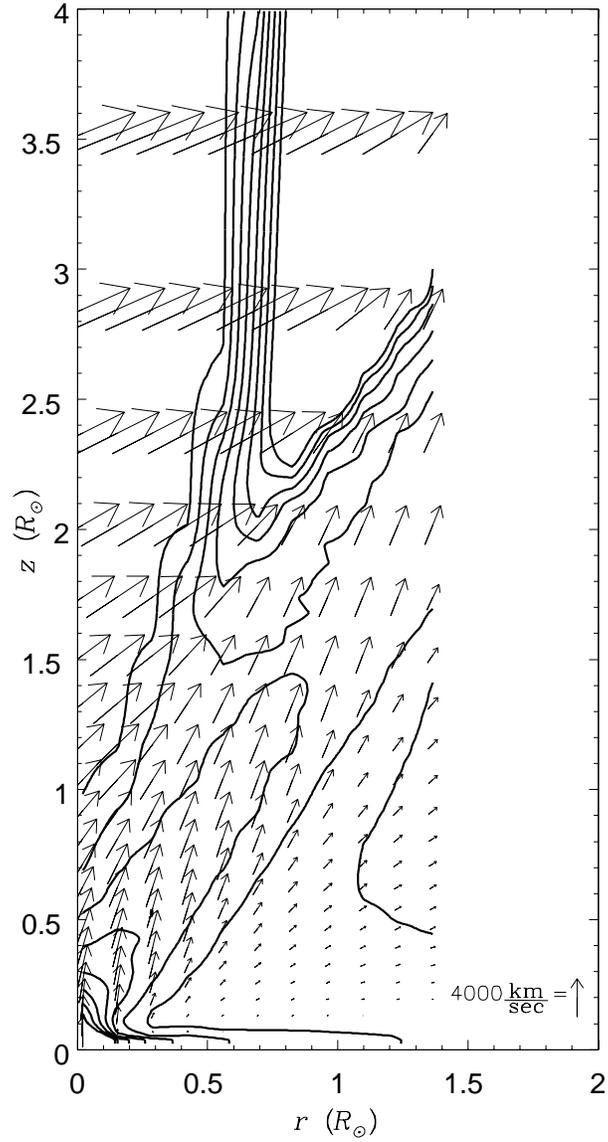}
\caption{
Vector field graph of wind velocity superimposed
with temperature contours for the model
applying local ionization equilibrium
and including radiative heating and cooling
(Model~H).
Parameters are similar to those of Figure~\ref{fig_mtf}.
}
\label{fig_mth}
\end{figure}

\begin{figure}
\epsscale{0.5}
\plotone{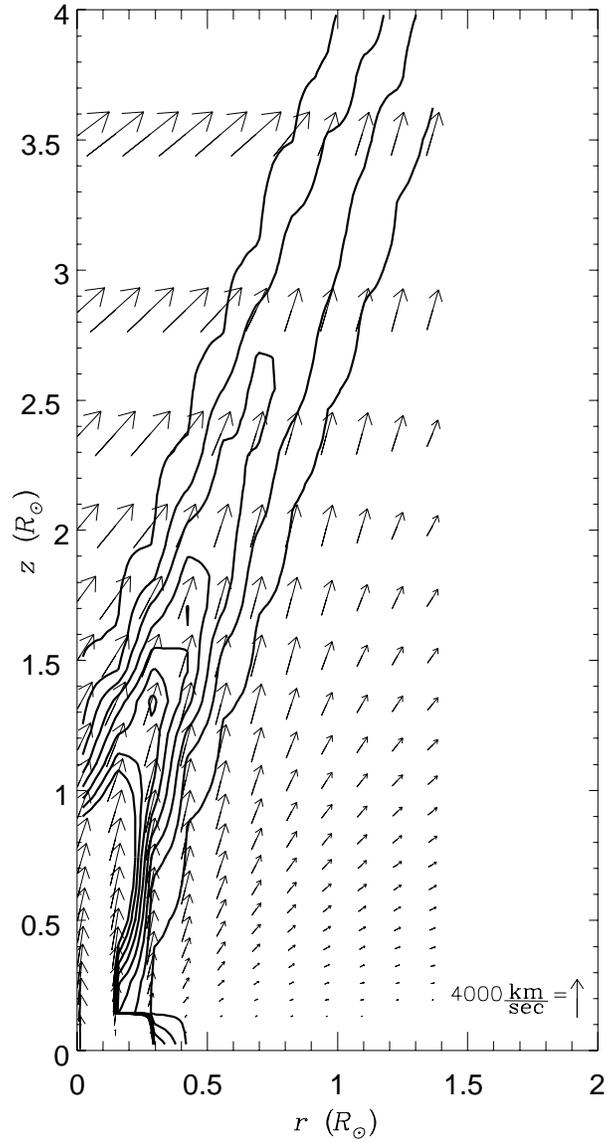}
\caption{
Vector field graph of wind velocity superimposed
with density contours for the model applying local
ionization equilibrium varying in space and time.
The computational grid starts at $r=R_{wd}$,
rather than $r=2 R_{wd}$ as in the results presented in
Figure~\ref{fig_mh}.
Parameters are similar to those of Figure~\ref{fig_mf}.
}
\label{fig_mc1}
\end{figure}

\begin{figure}
\epsscale{0.5}
\plotone{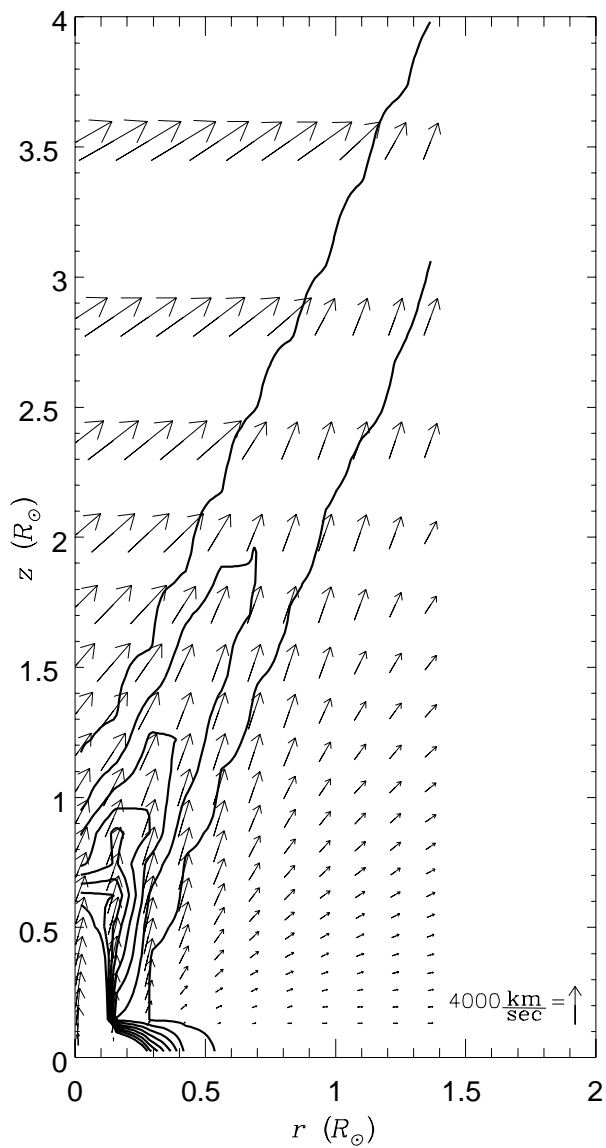}
\caption{
Vector field graph of wind velocity superimposed
with density contours for model disk wind applying local
ionization equilibrium varying in space and time.
The computational grid starts at $r=R_{wd}$,
rather than $r=2 R_{wd}$ as in the results presented in
Figure~\ref{fig_mh}.
The initial velocity set at the base of the wind is
$1 \, {\rm km \, s}^{-1}$,
rather than $10 \, {\rm km \, s}^{-1}$ as in the results presented
in Figure~\ref{fig_mh}.
Parameters are similar to those of Figure~\ref{fig_mf}.
}
\label{fig_mc2}
\end{figure}

\end{document}